\documentclass[preprint,showpacs,superscriptaddress,aps]{revtex4-1}
\usepackage{amssymb}
\usepackage{amsmath,amssymb,graphicx}
\usepackage{graphicx}
\usepackage{dcolumn}
\usepackage{bm}
\def\be{\begin{equation}}
\def\ee{\end{equation}}
\def\bea{\begin{eqnarray}}
\def\eea{\end{eqnarray}}

\def\lsim{\raise0.3ex\hbox{$\;<$\kern-0.75em\raise-1.1ex\hbox{$\sim\;$}}}
\def\gsim{\raise0.3ex\hbox{$\;>$\kern-0.75em\raise-1.1ex\hbox{$\sim\;$}}}

\usepackage{pstricks}
\begin{document}

\title{Observation of $CP$ violation in $D^0 \rightarrow K^- \pi^+ $  as a smoking gun for New Physics}

\author{David Delepine}
\email{delepine@fisica.ugto.mx}
\affiliation{{\fontsize{10}{10}\selectfont{Division de Ciencias e Ingenier\'ias,  Universidad de Guanajuato, C.P. 37150, Le\'on, Guanajuato, M\'exico.}}}

\author{Gaber Faisel}
\email{gfaisel@ncu.edu.tw}
\affiliation{{\fontsize{10}{10}\selectfont{Department of Physics
and Center for Mathematics and Theoretical Physics, National
Central University, Chung-Li, Tawian 32054.}}}
\affiliation{{\fontsize{10}{10}\selectfont{Egyptian Center for
Theoretical Physics, Modern University for Information and
Technology, Cairo, Egypt.}}}

\author{ Carlos A. Ramirez}
\email{jpjdramirez@yahoo.com}
\affiliation{{\fontsize{10}{10}\selectfont{Depto. de F\'isica,
Universidad de los Andes, A. A. 4976-12340, Bogot\'a, Colombia.}}}

\begin{center}

\begin{abstract}
In this paper, we study the  Cabibbo favored  non-leptonic $D^0$
decays into $ K^- \pi^+$ decays.  First we show that, within the
Standard Model, the corresponding CP asymmetry is strongly
suppressed and out of the experimental range even taking into
account the large strong phases coming from final state
Interactions.  We show also that although new physics models with
extra sequential generation can enhance the CP asymmetry by few
orders of magnitude however the resulting CP asymmetry is still
far from experimental range. The most sensitive New Physics Models
to this CP asymmetry comes from no-manifest Left-Right models
where a CP asymmetry up to 10\% can be reached and general two
Higgs models extension of SM where a CP asymmetry of order
$10^{-2}$ can be obtained without being in contradiction with the
experimental constraints on these models.
\end{abstract}
\end{center}

\maketitle
\section{Introduction}

The Standard Model (SM) has been very successful in predicting and
fitting all the experimental measurements  up-to-date  over
energies ranging many orders of magnitude\cite{Beringer:1900zz}.
Unfortunately the  SM is only a patchwork  where several sectors
remain totally unconnected. Flavor physics for example involves
quark  masses, mixings angles and CP violating phases appearing in
the Cabibbo-Kobayashi-Maskawa  (CKM) quarks mixing
matrix\cite{Cabibbo:1963yz,Kobayashi:1973fv}. These parameters
unavoidably have to be measured and are independent from
parameters present in other sectors like Electroweak Symmetry
breaking, Quantum Chromodynamics,  etc. Other sectors remain to be
tested like CP violation in the up-quarks sector and even tensions
with experimental measurements remain to be cleared (see for
instance
refs.\cite{Bona:2009cj,Hara:2010dk,Aubert:2009wt,Lees:2012xj}).

 This is why it is important to find processes where the SM
predictions are very well known and a simple measurement can show
their discrepancy.  One of these processes is the rare decays and
other \lq null' tests which correspond to an observable strictly
equal to zero within SM. So any deviation from zero of these \lq
null' tests observables is a clear signal of Physics beyond SM.
This is the case of Cabibbo-Favored (CF) and Double Cabibbo
Suppressed (DCS) non-leptonic charm decays where the direct CP
violation is very suppressed given that penguin diagrams are
absent\cite{Ryd:2009uf,Artuso:2008vf,Antonelli:2009ws}.

 Even with the observation of  $D^0$ oscillation
\cite{Aaltonen:2007ac,Staric:2007dt,Aubert:2007wf,:2012di,Lees:2012qh,Asner:2012xb}
and the first signal of CP violation in $D\to 2\pi,\ 2K$ (Singly
Cabibbo Suppressed (SCD) modes)
\cite{Frabetti:1993fw,Aitala:1996sh,Aitala:1997ff,Bonvicini:2000qm,Link:2001zj,Csorna:2001ww,Arms:2003ra,
Acosta:2004ts,Aubert:2006sh,Aubert:2007if,Staric:2008rx,Mendez:2009aa,Ko:2010ng,Aaij:2011in,Collaboration:2012qw,
Charles:2012rn}, it is not clear that the SM
\cite{Burdman:2001tf,Grossman:2006jg,Golowich:2006gq,Golowich:2007ka,Bigi:2011re,Hochberg:2011ru,
Brod:2011re,Cheng:2012wr,Giudice:2012qq} can describe correctly
the CP violation in the up quarks sector. It is even more
difficult as large distance contributions are important and
difficult to be evaluated
\cite{Donoghue:1985hh,Golowich:1998pz,Falk:2004wg,Cheng:2010rv,Gronau:2012kq}.
From the point of view of New Physics (NP), CP violation in CF and
DCS modes is an excellent opportunity given that it is very
suppressed in the SM and it is not easy to find a NP model able to
produce a reasonable CP violation signal. Thus measuring CP
violation in these channels is a very clear signal of New Physics.

Up to now, only  $D^0\leftrightarrow \bar D^0$ oscillations have
been observed and their parameters have been
measured\cite{Beringer:1900zz,Aaltonen:2007ac,Staric:2007dt,Aubert:2007wf,:2012di,Lees:2012qh,Asner:2012xb}:

\begin{eqnarray}
x\equiv {\Delta m_d\over \Gamma_D}=0.55^{+0.12}_{-0.13} \  \   & , & y\equiv {\Delta \Gamma_D\over 2\Gamma_D}=0.83(13)\\,
\left|{q\over p}\right|=0.91^{0.18}_{0.16} \  \ &,&\ \phi\equiv {\rm arg}\ (q/p) =-\left(10.2^{+9.4}_{-8.9}\right)^\circ
\end{eqnarray}
where $x\neq 0$ or/and $y\neq 0$ mean oscillations have been
observed, while $|q/p|\neq 1$ and/or $\phi\neq 0$ are necessary to
have CP violation. The theoretical estimations of these
parameters\cite{Beringer:1900zz} are not easy as they  have large
uncertainties given that the $c$ quark is not heavy enough to
apply Heavy quark effective theory (HQE) (like in $B$
physics)\cite{Georgi:1992as}. Similarly it is not light enough to
use Chiral Perturbation Theory (CPTh) (like in Kaon physics).
Besides there are cancellations due to the GIM
mechanism\cite{Cabibbo:1963yz,Glashow:1970gm}. Theoretically CP
violation  in the charm sector is smaller than in the $B$ and kaon
sectors. This is due to a combination factors: CKM matrix elements
($\left|V_{ub}V_{cb}^*/V_{us}V_{cs}^*\right|^2\sim 10^{-6}$) and
the fact that $b$ quark mass is  small compared to top mass. CP
violation  in the $b$-quark sector is due to the large top quark
mass, while in the kaon is due to a combination of the charm and
top quark.

Experimental data  should be  improved within the next years with
LHCB \cite{Gersabeck:2011zz} and the different Charm Factory
project \cite{Aushev:2010bq}. In table (\ref{table1}) the
experimentally measured  Branching ratios and CP asymmetries are
given for different non-leptonic $D$ decays.

\begin{table}
  \centering
  \begin{tabular}{|l|l|l||l|l|l|}  \hline
    Mode & BR[\%]     &$A_{\rm CP}$ [\%]  &     Mode & BR[\%]     &$A_{\rm CP}$ [\%] \\    \hline
   $D^0 \to K^-\pi^+$\ CF           &3.95(5)     & -  &
   $D^0 \to \bar K^0\pi^0$\ CF      &2.4(1)      & -  \\     \hline
   $D^0 \to \bar K^0\eta$\ CF       &0.96(6)     & -  &
   $D^0 \to \bar K^0\eta'$\ CF      &1.90(11)    & -  \\     \hline
   $D^+ \to \bar K^0\pi^+$\ CF      &3.07(10)   & -   &
   $D_s^+ \to K^+\bar K^0$\ CF      &2.98(8)    & -  \\     \hline
   $D_s^+ \to \pi^+\eta$\ CF        &1.84(15)   & -   &
   $D_s^+ \to \pi^+\eta'$\ CF       &3.95(34)   & -  \\     \hline \hline

   $D^0 \to K^+\pi^-$\ DCS           &$1.48(7)\cdot 10^{-4}$   & -  &
   $D^0 \to K^0\pi^0$\ DCS           &-   & -  \\     \hline
   $D^0 \to K^0\eta$\ DCS           &-   & -   &
   $D^0 \to K^0\eta'$\ DCS           &-   & -  \\     \hline
   $D^+ \to K^0\pi^+$\ DCS           &-   & -  &
   $D^+ \to K^+\pi^0$\ DCS           &$1.72(19)\cdot 10^{-2}$   & -  \\     \hline
   $D^+ \to K^+\eta$\ DCS            &$1.08(17)\cdot 10^{-2}$   & -  &
   $D^+ \to K^+\eta'$\ DCS           &$1.76(22)\cdot 10^{-2}$   & -  \\     \hline
   $D_s^+ \to K^+K^0$\ DCS           &-   & -  &&& \\     \hline\hline

   $D^0 \to \pi^-\pi^+$             &0.143(3)   & 0.22(24)(11)  \\     \hline
   $D^0 \to K^- K^+$                &0.398(7)   & -0.24(22)(9)  &
 $A_{\rm CP}(K^+K^-)-A_{\rm CP}(\pi^+\pi^-)$ & -- & -0.65(18)  \\     \hline
   $D^+ \to K_S^0\pi^+$             &1.47(7)    & -0.71(19)(20)  &
   $D^\pm \to \pi^+\pi^- \pi^\pm$   &0.327(22)  & 1.7(42)  \\     \hline
   $D^\pm \to K^\mp\pi^\pm \pi^\pm$ &9.51(34)   & -0.5(4)(9)    &
   $D^\pm \to K_s^0\pi^\pm \pi^0$   &6.90(32)   & 0.3(9)(3)  \\     \hline
   $D^\pm \to K^+K^- \pi^\pm$       &0.98(4)    & 0.39(61)     &  & & \\ \hline

  \end{tabular}
  \caption{Direct CP in $D$ non-leptonic decays, from Heavy Flavor Averaging Group HAFG \cite{hafg,Beringer:1900zz}}
  \label{table1}
\end{table}

In this paper, we study in details the CP  asymmetry  for the CF
$D^0 \rightarrow  K^- \pi^+$ decay. In sect. II, we give the
general description of the Effective Hamiltonian describing this
decay within SM and show how to evaluate the strong phases needed
to get CP violating observables. These strong phases are generated
through Final State Interaction (FSI). In sect. III, we first
evaluate the SM prediction for the CP asymmetry and we show that
within SM, such CP asymmetry is  experimentally out of range. In
sect. IV,  New Physics models are introduced and their
contributions to CP asymmetry are evaluated. Finally, we conclude
in sect. V.

\section{ General Description of CF non leptonic $D^0$ decays into $K^-$ and $\pi^+$}

In general the Hamiltonian describing   $D^0 \to K^- \pi^+$  is given by

\begin{eqnarray}
{\cal L}_{\rm eff.} &=& {G_F\over\sqrt{2}}V_{cs}^*V_{ud} \left[\sum_{i,\ a} c_{1ab}^i \bar s\Gamma^ic_a\bar u \Gamma_id_b+ \sum_{i,\ a} c_{2ab}^i \bar u\Gamma^ic_a\bar s \Gamma_id_b \right]
\end{eqnarray}
with $i=$S, V and T for respectively scalar (S), vectorial (V) and
tensorial (T) operators. The Latin indexes $a,\ b=L,\ R$ and
$q_{L,\ R}=(1\mp \gamma_5)q$.

Within the SM, only two operators contribute to  the effective
hamiltonian for this
process\cite{Ryd:2009uf,Artuso:2008vf,Antonelli:2009ws}. The other
operators can only be generated through new physics.
\begin{eqnarray}
{\cal H} &=& {G_F\over\sqrt{2}}V_{cs}^*V_{ud}\left(c_1\bar s\gamma_\mu c_L\bar u \gamma^\mu d_L+c_2\bar u \gamma_\mu c_L\bar s\gamma^\mu d_L\right)+{\rm h.c.} \\
&=& {G_F\over\sqrt{2}}V_{cs}^*V_{ud}\left(c_1{\cal O}_1+c_2{\cal O}_2\right)+{\rm h.c.}
\label{SMH}
\end{eqnarray}
where $a_1\equiv c_1+ c_2/N_c =1.2\pm 0.1$ and $a_2\equiv
c_2-c_1/N_C=-0.5\pm
0.1$\cite{Ryd:2009uf,Artuso:2008vf,Antonelli:2009ws}  where $N_C$
is the color number. For the case $D\to
K\pi$\cite{Ryd:2009uf,Artuso:2008vf,Antonelli:2009ws} one has that
\begin{eqnarray}
A_{D^0\to K^-\pi^+} &=&-i{G_F\over \sqrt{2}}V_{cs}^*V_{ud} \left[a_1X^{\pi^+}_{D^0K^-}+a_2X^{D^0}_{K^-\pi^+} \right], \label{1}\\
 BR&=&{\tau_D p_K\over 8\pi m_D^2}|A|^2
\end{eqnarray}
where BR is the Branching ratio of the process. $\tau_D$ is the D
lifetime, $p_K$ is the Kaon momentum and $m_D$ is the $D$ meson
mass. The $X^{\pi^+}_{D^0K^-}$ and $X^{D^0}_{K^-\pi^+}$ can be
expressed in the following way:
\begin{eqnarray}
X^{P_1}_{P_2P_3}= if_{P_1}\Delta_{P_2P_3}^2 F_0^{P_2P_3}(m_{P_1}^2),\  \Delta_{P_2P_3}^2=m_{P_2}^2-m_{P_3}^2
\end{eqnarray}
where $f_{D}$ and $f_{K}$ are  the  decay constants for $D$ and
$K$ mesons respectively and $F_0^{DK}$ and $F_0^{D\pi}$ are the
corresponding form factors. These amplitudes have been computed
within the so called naive factorization approximation (NFA)
without including the Final State Interaction (FSI). In NFA, no
strong CP conserving phases are obtained (and therefore no CPV is
predicted) but it is well known that FSI effects are very
important in these channels
\cite{Buccella:1994nf,Falk:1999ts,Rosner:1999xd,Gao:2006nb,Cheng:2010ry}.
In principle you have many FSI contributions: resonances, other
intermediate states, rescattering, and so on. Resonances are
specially important in this region given that they are abundant.
They can be included and seems to produce appropriate strong
phases \cite{Cheng:2010ry}. However the other contributions
mentioned above have to be included too, rendering the theoretical
prediction cumbersome. A more practical approach, although less
predictive, is obtained by fitting the experimental data
\cite{Buccella:1994nf,Cheng:2010ry}. This is the so called quark
diagram approach.  Within this approach, the amplitude is
decomposed into  parts corresponding to generic quark diagrams.
The main contributions are the tree level quark contribution (T),
exchange quark diagrams (E), color-suppressed quarks diagrams (C).
Their results can be summarized in the following way, for the
process under consideration\cite{Cheng:2010ry}:
\begin{eqnarray}
A_{D^0\to K^-\pi^+} &\equiv &  V_{cs}^*V_{ud}(T+E) \label{4}
\end{eqnarray}
with \begin{eqnarray}
T &=&  (3.14\pm 0.06)\cdot 10^{-6}{\rm GeV}\nonumber \\
E &=&  1.53^{+0.07}_{-0.08}\cdot 10^{-6}\cdot {\rm e}^{(122\pm 2)^\circ\ i}\ {\rm GeV} \label{amplitude}
\end{eqnarray}
where in NFA  they can be approximately written  as
\begin{eqnarray}
T& \simeq & {G_F\over \sqrt{2}}a_1f_\pi(m_D^2-m_K^2)F_0^{DK}(m_\pi^2) \\
E& \simeq & -{G_F\over \sqrt{2}}a_2f_D(m_K^2-m_\pi^2)F_0^{K\pi}(m_D^2)
\end{eqnarray}
In the rest of this work we are going to use the values obtained
by the experimental fit, given in eq. (\ref{amplitude}).

\section{ CP asymmetry in $D^0 \to K^- \pi^+$ within SM}

\begin{figure}[tbhp]
  \includegraphics[width=4.5cm]{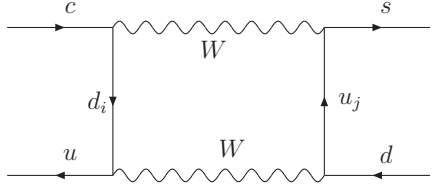}\hspace{1.cm}

  \caption{Feynman diagram for CF processes: Box  contribution.}
\end{figure}

In the case of CF (and DCF) processes the corrections are very
small (see diagrams in fig.1 and fig.2) and are generated through
box and di-penguin
diagrams\cite{Donoghue:1986cj,Petrov:1997fw,box}. In this section,
we shall evaluate these contributions.

The box contribution is given as \cite{box,He:2009rz}

\begin{eqnarray}
\Delta {\cal H} &=& \frac{G_F^2m_W^2}{ 2\pi^2} V_{cD}^*V_{uD} V_{Us}^*V_{Ud}f(x_U,\ x_D)  \bar u\gamma_\mu c_L\bar s\gamma^\mu d_L \\
&=&\frac{G_F^2m_W^2}{ 2\pi^2} \lambda^D_{cu}\lambda^U_{sd}f(x_U,\ x_D){\cal O}_2 \\
&=&\frac{G_F^2m_W^2}{ 2\pi^2}  b_x {\cal O}_2\nonumber
\end{eqnarray}
where
\begin{eqnarray}
b_x &\equiv & \lambda^D_{cu}\lambda^U_{sd}f(x_U,\ x_D) \\
 &=&  V_{cd}^*V_{ud}\left( V_{us}^*V_{ud}f_{ud}+V_{cs}^*V_{cd}f_{cd}+V_{ts}^*V_{td}f_{td}  \right)
\nonumber \\
&& + V_{cs}^*V_{us}\left( V_{us}^*V_{ud}f_{us}+V_{cs}^*V_{cd}f_{cs}+V_{ts}^*V_{td}f_{ts}  \right)  +V_{cb}^*V_{ub}\left( V_{us}^*V_{ud}f_{ub}+V_{cs}^*V_{cd}f_{cb}+V_{ts}^*V_{td}f_{tb}  \right)
\nonumber \\
&=& V_{cs}^*V_{us}\left[ V_{cs}^*V_{cd}\left(f_{cs}-f_{cd}-f_{us}+f_{ud}\right)+V_{ts}^*V_{td}\left(f_{ts}-f_{td}  -f_{us}+f_{ud} \right) \right] \nonumber \\
& & + V_{cb}^*V_{ub}\left[V_{cs}^*V_{cd}\left(f_{cb}-f_{cd} -f_{ub}+ f_{ud}\right)+V_{ts}^*V_{td}\left(f_{tb}-f_{td}-f_{ub}+f_{ud} \right) \right]
\end{eqnarray}
with $\lambda_{DD'}^U \equiv V_{UD}^*V_{UD'}$, $\lambda_{UU'}^D \equiv V_{UD}^*V_{U'D}$,  $U=u,\ c,\ t$ and $D=d,\ s,\ b$,  $x_q=(m_q/m_W)^2$ and $f_{UD} \equiv f(x_U,x_D)$ \cite{inami}
\begin{eqnarray}
f(x,\ y) ={7xy-4\over 4(1-x)(1-y)} +{1\over x-y}\left[ {y^2\log y\over (1-y)^2}\left(1-2x+{xy\over 4}\right)-  {x^2\log x\over (1-x)^2}\left(1-2y+{xy\over 4}\right)   \right] \nonumber
\end{eqnarray}
Numerically, one obtains
\begin{equation}
b_x \simeq 3.6\cdot 10^{-7} {\rm e}^{0.07\cdot i}
\end{equation}
The quark masses are taken their values at $m_c$ scale as given in \cite{Beringer:1900zz}.
The other contribution to the Lagrangian is the dipenguin and it gives \cite{Donoghue:1986cj,Petrov:1997fw,Chia:1983hd}

\begin{figure}[tbhp]
  \includegraphics[width=7.5cm]{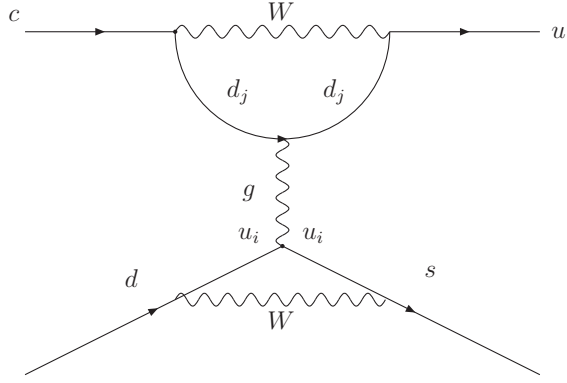}

  \caption{Feynman diagram for CF processes:di-penguins contribution.}
\end{figure}

\begin{eqnarray}
\Delta {\cal H} &=& -{G_F^2\alpha_S\over 8\pi^3}\left[  \lambda^D_{cu}  E_0(x_D)\right] \left[ \lambda^U_{sd} E_0(x_U)\right]    \bar s \gamma_\mu T^a d_L \left(g^{\mu \nu} \Box-\partial^\mu\partial^\nu \right) \bar u \gamma_\nu T^a c_L\nonumber \\
&=&  -{G_F^2\alpha_S\over 8\pi^3}pg \bar s \gamma_\mu T^a d_L \left(g^{\mu \nu} \Box-\partial^\mu\partial^\nu \right) \bar u \gamma_\nu T^a c_L \\
&\equiv & {G_F^2\alpha_S\over  16\pi^3} p_g {\cal O} \nonumber \\
p_g&\equiv & \left[  \lambda^D_{cu}  E_0(x_D)\right] \left[ \lambda^U_{sd} E_0(x_U)\right]=\left[ V_{cs}^*V_{us}\left(E_0(x_s)-E_0(x_d)\right)+ V_{cb}^* V_{ub}\left( E_0(x_b)-E_0(x_d) \right)   \right]
\nonumber \\
 && \left[ V_{cd}V_{cs}^*  \left(E_0(x_c)-E_0(x_u)\right) + V_{td}V_{ts}^* \left(E_0(x_t)-E_0(x_u)\right)\right]
\end{eqnarray} where $T^a$ are the generator of $SU(3)_C$.
Numerically, $p_g \simeq -1.62\cdot{\rm e}^{-0.002 i}$ and the Inami functions are given by
\begin{eqnarray}
E_0(x) &=&     {1\over 12(1-x)^4}\left[ x(1-x)(18-11x-x^2)-2(4-16x+9x^2)\log(x)\right]
\end{eqnarray}
The operator ${\cal O}$ can be reduced as
\begin{eqnarray}
{\cal O} &=& \bar s \gamma_\mu T^a d_L \left(g^{\mu \nu} \Box-\partial^\mu\partial^\nu \right) \bar u \gamma_\nu T^a c_L  = \bar s \gamma_\mu T^a d_L \Box \left(\bar u \gamma^\nu T^a c_L\right) + \bar s \partial \hskip-0.2cm /\ T^a d_L  \bar u \partial \hskip-0.2cm /\ T^a c_L
\nonumber \\
&=& -q^2 \bar s \gamma_\mu T^a d_L \bar u \gamma^\mu T^a c_L-\left(m_s \bar s T^a d_{S-P}+ m_d\bar s T^a d_{S+P} \right) \cdot \left(m_c \bar u T^a c_{S+P}+m_u \bar u T^a c_{S-P}\right) \nonumber \\
 &&-q^2 \bar s \gamma_\mu T^a d_L \bar u \gamma^\mu T^a c_L -m_s m_c \bar s T^a d_L \bar u T^a c_R - m_d m_u\bar s T^a d_R \bar u T^a c_L \nonumber \\
 && -m_s m_u\bar s T^a d_L \bar u T^a c_L-m_d m_c\bar s T^a d_R \bar u T^a c_R
\end{eqnarray}
where $q^2$ is the gluon momentum and $N$ is the colour number.  This expression can be simplified using the fact that
\begin{eqnarray}
\bar s \gamma_\mu T^a d_L \bar u \gamma^\mu T^a c_L &=& {1\over 2}\left({\cal O}_1-{1\over N}{\cal O}_2\right) \nonumber \\
\bar s T^a d_L \bar u T^a c_R &=&-{1\over 4}\bar s \gamma_\mu c_R \bar u \gamma^\mu d_L -{1\over 2N}\bar s d_L \bar u c_R   \nonumber \\
\bar s T^a d_R \bar u T^a c_L &=& -{1\over 4}\bar s \gamma_\mu c_L \bar u \gamma^\mu d_R -{1\over 2N}\bar s d_R \bar u c_L \nonumber \\
\bar s T^a d_L \bar u T^a c_L &=&-{1\over 4}\bar s c_L \bar u d_L -{1\over 16}\bar s\sigma_{\mu\nu} c_L \bar u \sigma^{\mu\nu} d_L  -{1\over 2N}\bar s d_L \bar u c_L \nonumber \\
\bar s T^a d_R \bar u T^a c_R &=&-{1\over 4}\bar s c_R \bar u d_R -{1\over 16}\bar s\sigma_{\mu\nu} c_R \bar u \sigma^{\mu\nu} d_R-{1\over 2N}\bar s d_R \bar u c_R
\end{eqnarray}
Once taking the expectation values, one obtains
\begin{eqnarray}
\left<{\cal O}\right> &=&-q^2 \left<\bar s \gamma_\mu T^a d_L \bar u \gamma^\mu T^a c_L\right> -m_s m_c \left<\bar s T^a d_L \bar u T^a c_R\right>
-m_d m_u\left<\bar s T^a d_R \bar u T^a c_L\right> \nonumber \\
&&-m_s m_u\left<\bar s T^a d_L \bar u T^a c_L\right>-m_d m_c\left<\bar s T^a d_R \bar u T^a c_R\right>
\nonumber \\
&\simeq &-{q^2\over 2}\left(1-{1\over N^2}\right)X^{\pi^+}_{D^0K^-}+{m_sm_c\over 4}\left(1-{1\over N}\right)X^{\pi^+}_{D^0K^-}+{5m_d\over 8Nm_s}m_D^2X^{D^0}_{K^- \pi^+}
\end{eqnarray}
Hence, one gets for the Wilson coefficients

\begin{eqnarray}
\Delta a_1 &=& -{G_Fm_W^2\over \sqrt{2}\ \pi^2V_{cs}^*V_{ud}N } b_x- {G_F \alpha_S\over 4\sqrt{2} \pi^3V_{cs}V_{us}^*}\left[{q^2\over 2}\left(1-{1\over N^2}\right)-{m_cm_s\over 4}\left(1-{1\over N}\right)  \right]  p_g \nonumber \\
&\simeq & 2.8\cdot 10^{-8}{\rm e}^{-0.004i} \nonumber \\
\Delta a_2 &=& -{G_Fm_W^2\over \sqrt{2}\ \pi^2V_{cs}^*V_{ud} } b_x- {G_F \alpha_S\over 4\sqrt{2} \pi^3V_{cs}V_{us}^*}{5m_d m_D^2\over 8Nm_s}p_g \nonumber  \\
& \simeq & -2.0\cdot 10^{-9}{\rm e}^{0.07i}
\end{eqnarray}
where to obtain the last result it has been used the fact that for
the decay $D^0\to K^-\pi^+$, one can approximate $q^2=(p_c \mp
p_u)^2=(p_s\pm  p_d)^2\simeq
(p_D-p_\pi/2)^2=(m_D^2+m_K^2)/2+3m_\pi^2/4$, by assuming that
$p_c\simeq p_D$ and $p_u\simeq p_\pi/2$ and $\alpha_S\simeq 0.3$.
It should be noticed that the box contribution is dominated by the
heavy quarks while the penguin is by the light ones. The direct CP
asymmetry is then
\begin{eqnarray}
A_{CP} &=& {|A|^2-|\bar A|^2 \over |A|^2+|\bar A|^2 }= {2|r|\sin(\phi_2-\phi_1)\sin(\alpha_E)))\over |1+ r|^2 }=  1.4\cdot 10^{-10}
\end{eqnarray}
with $r=E/T$, $a_i\to a_i+\Delta a_i=a_i+|\Delta a_i|\exp[i\Delta
\phi_i]$ and $\phi_i\simeq\Delta a_i \sin\Delta \phi_i/a_i$ and
$\alpha_E$ is the conserving phase which appears in
eq.(\ref{amplitude}).

\section{New Physics}

With New Physics, the general Hamiltonian is not only given by
${\cal O}_{1,2}$.  The expressions of the expectation values of
these operators can be found in the appendix. It is important to
notice that as expected only two form factors appear, namely
$\chi^{D^0}_{K^- \pi^+}$ and $\chi^{\pi^+}_{D^0 K^-}$. This is
important to take into account the FSI interactions as the first
one is identified as E contribution and the second one is
identified as T contribution. In the next subsections, we shall
calculate the Wilson coefficient for different models of New
Physics. The first case will be assuming to have extra SM fermion
family. The second example will be to compute the CP asymmetry
generated by a new charged gauge boson as it appears for instance
in models based on gauge group $SU(2)_L \times SU(2)_R \times
U(1)_{B-L}$ and our last subsection is dedicated to the effects CP
asymmetry coming from new charged Higgs-like scalar fields,
applying to two Higgs extension of the SM (type II and type III).

\subsection{Contributions to $A_{CP}$ from extra SM fermion family}

A simple extension of the SM is the introduction of a new
sequential generation of quarks and leptons (SM4). A fourth
generation is not exclude by precision
data\cite{Frampton:1999xi,Maltoni:1999ta, He:2001tp,
Novikov:2002tk, Kribs:2007nz, Hung:2007ak,
Bobrowski:2009ng,Hashimoto:2010at}. Recent reviews  on
consequences of a fourth generation can be found in
\cite{Holdom:1986rn,Hill:1990ge,
Carpenter:1989ij,Hung:1997zj,Ham:2004xh, Fok:2008yg,Hou:2008xd,
Kikukawa:2009mu, Hung:2009hy,Holdom:2009rf,Hung:2009ia}.

The $B\to K \pi$ CP asymmetries puzzles is easily solved by a fourth generation
\cite{Soni:2008bc,Hou:2005hd,Hou:2006jy} with a mass within the following range\cite{Soni:2008bc}:
\begin{eqnarray}
400\; \mathrm{ GeV} <& m_{u_4} & < 600\; \mathrm{ GeV} .
\end{eqnarray}
The value of SM4 parameters compatibles with the high precision LEP measurements
\cite{Maltoni:1999ta,He:2001tp,Novikov:2002tk,Bobrowski:2009ng} are
\begin{eqnarray}\label{eq:benchmarks}
m_{u_4} - m_{d_4}  &&\simeq
  \left( 1 + \frac{1}{5} \ln \frac{m_H}{115 \; \mathrm{GeV}}
                         \right) \times 50 \; \mathrm{GeV}  \\
|V_{u d_4}|,|V_{u_4 d}| &&\lsim 0.04
\end{eqnarray}
where $V$  is the CKM  quark  mixing matrix which is now a $4\times 4$ unitary matrix.
The direct search limits from LEPII and CDF \cite{Achard:2001qw,Lister:2008is,Aaltonen:2009nr} are given by:
\begin{eqnarray}\label{LEP-CDF}
m_{u_4} & >& 311\; \mathrm{GeV}\;   \\
m_{d_4} & >& 338\; \mathrm{GeV}.  \nonumber
\end{eqnarray}
Direct search by Atlas and CMS coll. have excluded $m_{d_4}<480$
GeV and $m_{q_4}<350$ GeV
\cite{Magnin:2009zz,Eberhardt:2012ck,Alok:2010zj}, above the tree
level unitarity limit, $m_{u_4}<\sqrt{4\pi/3}\ v\simeq 504$ GeV.
But SM4 is far to be completely understood. Most of the
experimental constraints are model-dependent. For instance it has
been shown in \cite{Geller:2012wx} that the bound on $m_{u_4}$
should be  relaxed up to $m_{u_4} > 350 GeV$ if the decay $u_4
\rightarrow ht$ dominates. The recent LHC results which observe an
excess in the $H \rightarrow \gamma \gamma$ corresponding to a
Higgs mass around 125 $GeV$ \cite{ATLAS:2012ae,Chatrchyan:2012tx}
seems to exclude the SM4 scenario \cite{Djouadi:2012ae} but this
results is based on the fact that once we include the next-to
leading order electroweak corrections, the rate
$\sigma(gg\rightarrow H)\times Br(H\rightarrow \gamma \gamma) $ is
suppressed by more than 50\% compared to the rate including only
the leading order corrections \cite{Georgi:1977gs,
Djouadi:1991tka,Denner:2011vt,
Djouadi:2012ae,Kuflik:2012ai,Eberhardt:2012sb}. This could be a
signal of a non-perturbative regime which in SM4 can be easily
reached at this scale due to the fourth generation strong Yukawa
couplings. Therefore, direct and model-independent searches for
fourth generation families at collider physics are still necessary
to completely exclude the SM4 scenario.

The CP asymmetry in model with a fourth family is easy to compute as the
 contributions come from the same diagrams in the  SM with just adding an extra
 $u_4\equiv  t'$ and $d_4\equiv b'$. Similarly in ref.\cite{Alok:2010zj},
 it has been found that new CKM matrix elements can be obtained (all consistent with zero and for $m_{b'}=600$ GeV) to be
\begin{eqnarray}
s_{14} &=&|V_{ub'}|=0.017(14),\ s_{24}={|V_{cb'}|\over c_{14}}={0.0084(62)\over c_{14}},\ s_{34}={|V_{tb'}|\over c_{14}c_{24}}= {0.07(8)\over c_{14}c_{24}} \nonumber \\
|V_{t'd}| &=& |V_{t's}| = 0.01(1),\ |V_{t'b}| = 0.07(8),\ |V_{t'b'}|=0.998(6),\ |V_{tb}|\geq 0.98 \nonumber \\
 \tan \theta_{12} &=& \left|{V_{us}\over V_{ud}}\right|,\ s_{13}={|V_{ub}| \over c_{14}},\ \delta_{13}=\gamma=68^\circ  \nonumber \\
|V_{cb}| &=& |c_{13}c_{24}s_{23}-u_{13}^*u_{14}u_{24}^*|\simeq c_{13}c_{24}s_{23} \label{ckm4}
\end{eqnarray}
The two remaining phases ($\phi_{14}$ and $\phi_{24}$) are
unbounded. Thus the absolute values of the CKM elements for the
three families remain almost unchanged but not their phases. From
these values one obtains
\begin{eqnarray}
s_{13} &=& 0.00415,\ s_{12}=0.225,\ s_{23}=0.04,\ s_{14}=0.016,\ s_{24}=0.006,\ s_{34}=0.04
\end{eqnarray}
For a 4th sequential family  the maxima value for the CP violation is obtained as
\begin{eqnarray}
A_{{\rm CP}} &\simeq&-1.1\cdot 10^{-7}  \nonumber \\
\end{eqnarray}
where one uses $|V_{ub'}| = 0.06,\ |V_{cb'}|=0.03,\ |V_{tb'}|=0.25,\ \phi_{14}=-2.9,\ \phi_{24}=1.3$

This maximal value is obtained when the parameters  mentioned
above are varied in a the range allowed by the experiential
constrains, according to eq. \ref{ckm4} in a 'three sigma' range.
The phases are varied in the whole range, from $-\pi$ to $\pi$.
Thus one can obtain an enhancement of thousand that may be large
but still very far from the experimental possibilities.

\subsection{A new charged gauge boson as Left Right models}

In this section, we shall look to see what could be the effect on
the CP asymmetry coming from a new charged gauge boson coupled to
quarks and leptons. As an example of such models, we  apply our
formalism to a well known extension of the Standard Model based on
extending the SM gauge group including a gauge $SU(2)_R$
\cite{Pati:1973rp,Mohapatra:1974hk,Mohapatra:1974gc,Senjanovic:1975rk,Senjanovic:1978ev}.
So now, our gauge group defining the electroweak interaction is
given by $SU(2)_L \times SU(2)_R \times U(1)_{B-L}$. This SM
extension has been extensively studied in previous works (see for
instance refs.
\cite{Beall:1981ze,Cocolicchio:1988ac,Langacker:1989xa,Cho:1993zb,Babu:1993hx}
) and their parameters have been strongly constrained by
experiments
\cite{Beringer:1900zz,Alexander:1997bv,Acosta:2002nu,Abazov:2006aj,Abazov:2008vj,Abazov:2011xs}.
Recently,  CMS \cite{Chatrchyan:2012meb,Chatrchyan:2012sc} and
ATLAS \cite{Aad:2011yg,Aad:2012ej} at LHC have  improved
the bound on the scale of the $W_R$ gauge boson mass
\cite{Maiezza:2010ic}. The new diagrams contributing to $D \to K
\pi$  are similar to the SM tree-level diagrams with $W_L $ is
replaced by a $W_R$. These diagrams contribute to the effective
Hamiltonian in the following way assuming no mixing between  $W_L$
and $W_R$ gauge bosons :

\begin{eqnarray}
{\cal H}_{\rm LR} &=& {G_F\over\sqrt{2}}\left({g_R m_W\over g_L m_{W_R}}\right)^2V_{Rcs}^*V_{Rud}\left(c_1'\bar s\gamma_\mu c_R\bar u \gamma^\mu d_R+c_2'\bar u \gamma_\mu c_R\bar s\gamma^\mu d_R\right)+{\rm H.C.} \nonumber \\
&=& {G_F\over\sqrt{2}}V_{cs}^*V_{ud}\left(c_1{\cal O}_1+c_2{\cal O}_2\right)+{\rm H.C.}
\end{eqnarray}
where $g_{L}$ and $g_{R}$ are the gauge $SU(2)_{L}$ and
$SU(2)_{R}$ couplings respectively. $m_W$ and $m_{W_R}$ are
 the $SU(2)_{L}$ and $SU(2)_{R}$ charged gauge boson masses respectively. $ V_R$
is the quark mixing matrix which appears in the right sector of
the lagrangian similar to the CKM quark mixing matrix. This new
contribution  can enhance the SM prediction for the CP asymmetry
but still it is suppressed due to the limit on $M_{W_R}$ which has
to be of order $2.3$ TeV \cite{Maiezza:2010ic} in case of
no-mixing Left right models.

 In refs.\cite{Chen:2012usa,Lee:2011kn} it has been shown
that the mixing between the left and the right gauge bosons can
strongly enhance any CP violation in the Charm and muon sector.
This LR mixing is restricted by deviation to non-unitarity of the
CKM quark mixing matrix. The results were that  the Left-Right
(LR) mixing angle called $\xi$ has to be smaller than
0.005\cite{Wolfenstein:1984ay} and right scale $M_R$ bigger than
2.5 TeV\cite{Maiezza:2010ic}. If the Left-Right is not manifest
(essentially that $g_R$ could be different from $g_L$ at
Unification scale), the limit on  $M_R$ scale is much less
restrictive and the right gauge bosons could be as light as $0.3$
TeV \cite{Olness:1984xb}.  In such a case, $\xi$ can be as large
as $0.02$ if  large CP violation phases in the right sector are
present \cite{Langacker:1989xa} still compatible with experimental
data \cite{Jang:2000rk,Badin:2007bv,Lee:2011kn}. Recently,
precision measurement of the muon decay parameters done by TWIST
collaboration \cite{MacDonald:2008xf,TWIST:2011aa} put model
independent limit on $\xi$ to be smaller than 0.03 (taking
$g_L=g_R$). Let's now  compute the effect of the LR mixing gauge
boson on our CP asymmetry. So first, one defines the charged
current mixing matrix\cite{Chen:2012usa}
\begin{eqnarray}
\left(\begin{array}{c}
  W_L \\
  W_R
\end{array}\right) =
\left(\begin{array}{cc}
  \cos \xi & -\sin \xi \\
{\rm e}^{i\omega}\sin \xi & {\rm e}^{i\omega}\cos \xi
\end{array}\right)
\left(\begin{array}{c}
  W_1 \\
  W_2
\end{array}\right)\simeq
 \left(\begin{array}{cc}
  1 & - \xi \\
{\rm e}^{i\omega}\xi & {\rm e}^{i\omega}
\end{array}\right)
\left(\begin{array}{c}
  W_1 \\
  W_2
\end{array}\right)
\end{eqnarray}
where $W_1$ and $W_2$ are the mass eigenstates and $\xi\sim 10^{-2}$.
Thus the charged currents interaction part become
\begin{eqnarray}
{\cal L} &\simeq & -{1\over \sqrt{2}} \bar U \gamma_\mu \left(g_LVP_L+g_R\xi \bar V^RP_R\right)DW_1^\dagger-
{1\over \sqrt{2}} \bar U \gamma_\mu \left(-g_L\xi VP_L+g_R\bar V^RP_R\right)DW_2^\dagger
\end{eqnarray}
where $V=V_{\rm CKM}$ and $\bar V^R={\rm e}^{i\omega}V^R$.
Once one integrates out the $W_1$ in the usual way and
neglecting the $W_2$ contributions given its mass is much higher,
one obtains the effective hamiltonian responsible of our process:
\begin{eqnarray}
{\cal H}_{\rm eff.} &=& {4G_F\over \sqrt{2}}\left[c_1 \  \bar{s}\gamma_\mu \left(V^*P_L+{g_R\over g_L}\xi \bar V^{R*}P_R  \right)_{cs}c \ \ \bar{u}\gamma^\mu  \left(VP_L+{g_R\over g_L}\xi \bar VP_R  \right)_{ud}  d \right.
\nonumber  \\
&& \left. c_2 \ \bar s_\alpha \gamma_\mu \left(V^*P_L+{g_R\over g_L}\xi \bar V^{R*}P_R  \right)_{cs}c_\beta\bar u_\beta\gamma^\mu  \left(VP_L+{g_R\over g_L}\xi \bar VP_R  \right)_{ud} d_\alpha
\right]+{\rm h.\ c.}
\nonumber \\
\end{eqnarray}
where $\alpha,\beta$ are color indices. It is easy to check that
taking the limit $\xi \to 0$, one obtains eq.(\ref{SMH}) with the
only difference comes from the $c_2$ terms, the Fierz transformation has
been  applied. The terms of the effective  Hamiltonian
proportional to $\xi$ are: \bea \Delta {\cal H}_{\rm eff}&\simeq &
{G_F\over \sqrt{2}}{g_R\over g_L}\xi \left[c_1\bar s
\gamma_\mu V_{cs}^*c_L\bar u \gamma^\mu
\bar V^R_{ud}d_R+ c_1 \bar s\gamma_\mu \bar
V^{R*}_{cs}c_R \bar u \gamma^\mu  V_{ud}d_L
\right.
\nonumber  \\
&& \left. c_2\bar s_\alpha \gamma_\mu V_{cs}^*c_{L\beta} \bar u_\beta \gamma^\mu
\bar V^R_{ud}d_{R\alpha}+ c_2 \bar s_\alpha \gamma_\mu \bar V^{R*}_{cs}c_{R\beta} \bar u_\beta \gamma^\mu  V_{ud}d_{L\alpha}
\right]+{\rm h.\ c.}
\eea
 The contribution to the amplitude  proportional to $\xi$ is then given by:
\begin{eqnarray}
 \Delta A &= & -{iG_F\over \sqrt{2}}{g_R\over g_L}\xi \left[-c_1 V_{cs}^*\bar V^R_{ud}
 \left( X^{\pi^+}_{D^0K^-}+{2\over N}\chi^{D^0} X^{D^0}_{K^-\pi^+} \right)+ c_1 \bar V^{R*}_{cs} V_{ud}  \left( X^{\pi}_{D^0K^-}+{2\over N}\chi^{D^0} X^{D^0}_{K^-\pi^+} \right) \right.
\nonumber  \\
&& \left. -c_2 V_{cs}^*\bar V^R_{ud} \left(2\chi^{D^0} X^{D^0}_{K^-\pi^+}+{1\over N} X^{\pi^+}_{D^0K^-} \right)+ c_2\bar V^{R*}_{cs}V_{ud} \left(2 \chi^{D^0} X^{D^0}_{K^-\pi^+}+{1\over N} X^{\pi^+}_{D^0K^-} \right)\right]
\nonumber  \\
&= & {iG_F\over \sqrt{2}}{g_R\over g_L}\xi \left(V_{cs}^*\bar V^R_{ud}-\bar V^{R*}_{cs} V_{ud}\right)\left( a_1 X^{\pi^+}_{D^0K^-}+ 2\chi^{D^0}  a_2X^{D^0}_{K^-\pi^+} \right) \nonumber \\
&=&-{g_R\over g_L}\xi \left(\bar V^{R*}_{cs} V_{ud}-V_{cs}^*\bar V^R_{ud}\right)\left( T- 2\chi^{D^0} E \right)
\end{eqnarray}
where$\chi^{\pi^+} $ and $\chi^{D^0}$ are defined as \bea \chi^{\pi^+}&=&{m_\pi^2\over (m_c-m_s)(m_u+m_d)}\nonumber\\
\chi^{D^0}&=& {m_D^2\over (m_c+m_u)(m_s-m_d)}\eea
The CP asymmetry becomes
\begin{eqnarray}
A_{\rm CP} ={4(g_R/g_L)\xi \over V_{cs}^*V_{ud}|1+r|^2}\left(1+2\chi^{D^0}\right)
 {\rm Im}\left(\bar V_{cs}^{R*}V_{ud}-V_{cs}^* \bar V_{ud}^R \right)
 {\rm Im}(r)
\end{eqnarray}
with $r=E/T$. For a value as large as  $\xi\sim 10^{-2}$ the
asymmetry can be as large as 0.1.  Also, we should notice that to
obtain this results, it has been  used the fact that the
chiralities don't mix under strong interactions, if the quark
masses are not taken into account. This is approximately the case
in the evolution of the Wilson coefficients from $m_W$ to $m_c$ as
the quark in the loop are the down quarks contrarily to process
like $b \to s \gamma$ where the quarks in the QCD corrections are
the up quarks and in that case, a strong effect from top quarks
could be expected
\cite{Cho:1991cj,Buras:1993xp,Chetyrkin:1996vx,Buras:2011we}. In
our case, as a first approximation, the QCD corrections to the
Wilson coefficient coming from the running of the renormalization
group from $m_W$ to $m_c$ can be safely neglected.
\subsection{Models with Charged Higgs contributions}\label{SMHeff}
Our last example of new physics is considering contribution to the
effective Hamiltonian responsible of the $D^0 \to K^- \pi^+$
process due to a new charged Higgs fields. The simple SM
extensions which include new charged Higgs fields are the two
Higgs doublet models (2HDM)\cite{Haber:1978jt,Abbott:1979dt}.
Usually, it is used to classify these 2HDM in three types: type I,
II or III (for a review see ref. \cite{Branco:2011iw}). In 2HDM
type II models (like Minimal Supersymmetric Standard Model), one
Higgs couples to the down quarks and charged leptons and the other
Higgs couples to up type quarks. LEP has performed a Direct search
for a charged Higgs in type II 2HDM  and they obtained a bound of
$78.6$ GeV \cite{Searches:2001ac}. Recent results on $ B\to  \tau
\nu$ obtained by BELLE \cite{Hara:2010dk} and BABAR
\cite{Aubert:2009wt}  have strongly improved the indirect
constraints on the charged Higgs mass in type II 2HDM
\cite{Baak:2011ze}:
\begin{equation}
m_{H^+}> 240 GeV \ \  at \ \ 95 \% CL
\end{equation}
2HDM type III is a general model where both Higgs couples to up
and down quarks. Of course, this means that 2HDM type III can
induce Flavor violation in Neutral Current and thus  it can be
used to  strongly constrain the new parameters in the model. We
shall focus our interest to the two Higgs doublet of type III as
the other two can be obtained from type III taking some limits. In
the 2HDM of type III, the Yukawa Lagrangian can be written as
\cite{Crivellin:2010er,Crivellin:2012ye} :
\begin{eqnarray}
\mathcal{L}^{eff}_Y &=& \bar{Q}^a_{f\,L} \left[
  Y^{d}_{fi} \epsilon_{ab}H^{b\star}_d\,-\,\epsilon^{d}_{fi} H^{a}_u \right]d_{i\,R}\\
&-&\bar{Q}^a_{f\,L} \left[ Y^{u}_{fi}
 \epsilon_{ab} H^{b\star}_u \,+\, \epsilon^{ u}_{fi} H^{a}_d
  \right]u_{i\,R}\,+\,\rm{H.c}. \,,\nonumber
\end{eqnarray}
where $\epsilon_{ab}$ is the totally antisymmetric tensor, and
$\epsilon^q_{ij}$ parametrizes the non-holomorphic corrections
which couple up (down) quarks to the down (up) type Higgs doublet.
After electroweak symmetry breaking, $\mathcal{L}^{eff}_Y$ gives
rise to   the following charged Higss-quarks interaction
Lagrangian:
\begin{equation}
\mathcal{L}^{eff}_{H^\pm} = \bar{u}_f {\Gamma_{u_f d_i
}^{H^\pm\,LR\,\rm{eff} } }P_R d_i
+ \bar{u}_f {\Gamma_{u_f d_i }^{H^\pm\,RL\,\rm{eff} } }P_L d_i\, ,\\
 \label{Higgs-vertex}
\end{equation}
with \cite{Crivellin:2012ye} \bea {\Gamma_{u_f d_i
}^{H^\pm\,LR\,\rm{eff} } } &=& \sum\limits_{j = 1}^3 {\sin\beta\,
V_{fj} \left( \frac{m_{d_i }}{v_d} \delta_{ji}-
  \epsilon^{ d}_{ji}\tan\beta \right), }
\nonumber\\
{\Gamma_{u_f d_i }^{H^ \pm\,RL\,\rm{eff} } } &=& \sum\limits_{j =
1}^3 {\cos\beta\,  \left( \frac{m_{u_f }}{v_u} \delta_{jf}-
  \epsilon^{ u\star}_{jf}\tan\beta \right)V_{ji}}
 \label{Higgsv}
\eea
Here $v_u$ and $v_d$ are the vacuum expectations values of  the
neutral component of the  Higgs doublets,  $V$ is the CKM matrix and $tan \beta = v_u/v_d$.
Using the Feynman-rule given in Eq. (\ref{Higgs-vertex}) we can
compute the effective Hamiltonian resulting from the tree level
exchanging charged Higgs diagram that governs the process under
consideration namely,
\be {\mathcal H}_{eff}= \frac{  G_F}{\sqrt{2}}V^*_{cs}V_{ud}
\sum^4_{i=1} C^H_i(\mu) Q^H_i(\mu),\ee where $C^H_i$ are the
Wilson coefficients obtained by perturbative QCD running from
$M_{H^{\pm}}$ scale to the scale $\mu$ relevant for hadronic decay
and $Q^H_i$ are the relevant local operators at low energy scale
$\mu\simeq m_c$. The operators can be written as %
\bea
Q^H_1 &=&(\bar{s} P_R c)(\bar{u} P_L d),\nonumber\\
Q^H_2 &=&(\bar{s} P_L c)(\bar{u} P_R d),\nonumber\\
Q^H_3 &=&(\bar{s} P_L c)(\bar{u} P_L d),\nonumber\\
Q^H_4 &=&(\bar{s} P_R c)(\bar{u} P_R d),
 \eea
And the Wilson coefficients $C^H_i$, at the electroweak scale,
are given by
\begin{eqnarray}
C^H_1 &=& \frac {\sqrt{2} }{ G_F V^*_{cs}V_{ud}
 m^2_H} \bigg(\sum\limits_{j = 1}^3
{\cos\beta\, V_{j1} \left( \frac{m_u }{v_u} \delta_{j1}-
\epsilon^{ u\star}_{j1}\tan\beta \right)}\bigg)\bigg(
\sum\limits_{k= 1}^3 {\cos\beta\,V^{\star}_{k2}} \left(
\frac{m_c}{v_u} \delta_{k2}-\epsilon^{ u}_{k2}\tan\beta
\right)\bigg),\nonumber\\
C^H_2 &=& \frac {\sqrt{2} }{ G_F V^*_{cs}V_{ud}
 m^2_H} \bigg(\sum\limits_{j = 1}^3
{\sin\beta\,V_{1j}  \left( \frac{m_d }{v_d} \delta_{j1}-
\epsilon^{ d}_{j1}\tan\beta \right)}\bigg)\bigg( \sum\limits_{k=
1}^3 {\sin\beta\,V^{\star}_{2k}} \left( \frac{m_s}{v_d}
\delta_{k2}-\epsilon^{ d\star}_{k2}\tan\beta
\right)\bigg)\nonumber\\
 C^H_3 &=& \frac {\sqrt{2} }{  G_F V^*_{cs}V_{ud}
 m^2_H} \bigg(\sum\limits_{j = 1}^3 {\cos\beta\, V_{j1} \left(
\frac{m_u }{v_u} \delta_{j1}- \epsilon^{ u\star}_{j1}\tan\beta
\right)}\bigg)\bigg( \sum\limits_{k= 1}^3
{\sin\beta\,V^{\star}_{2k}} \left( \frac{m_s}{v_d}
\delta_{k2}-\epsilon^{ d\star}_{k2}\tan\beta
\right)\bigg),\nonumber\\
C^H_4 &=& \frac {\sqrt{2} }{ G_F V^*_{cs}V_{ud}
 m^2_H}\bigg( \sum\limits_{k= 1}^3
{\cos\beta\,V^{\star}_{k2}} \left( \frac{m_c}{v_u}
\delta_{k2}-\epsilon^{ u}_{k2}\tan\beta
\right)\bigg)\bigg(\sum\limits_{j = 1}^3 {\sin\beta\,V_{1j} \left(
\frac{m_d }{v_d} \delta_{j1}- \epsilon^{ d}_{j1}\tan\beta
\right)}\bigg) \nonumber \\
 \label{Higgsw}
\end{eqnarray}
We now discuss the experimental constraints on the $ \epsilon^{
q}_{ij}$ where $q=d,u$. The  flavor-changing elements
$\epsilon^d_{ij}$ for $i\neq j$ are strongly  constrained from
FCNC processes in the down sector because of tree-level neutral
Higgs exchange. Thus, we are left with only
$\epsilon^d_{11},\epsilon^d_{22}$. Concerning the elements
$\epsilon^u_{ij}$ we see that only
$\epsilon^u_{11},\epsilon^u_{22}$ can significantly effects the
Wilson coefficients without any CKM suppression. Other
$\epsilon^u_{ij}$ terms will be so small as the CKM suppression
will be of  orders  $\lambda$ or $\lambda^2$ or higher and so we
neglect them in our analysis.
One of the important constraints that on $ \epsilon^{
q}_{ij}$ where $q=d,u$ can be obtained by applying
 the naturalness criterion of 't Hooft to the quark masses.
According to  the naturalness criterion of 't Hooft,  the
smallness of a quantity is only natural if a symmetry is gained in
the limit in which this quantity is zero \cite{Crivellin:2012ye}.
Thus it is unnatural to have large accidental cancellations
without a symmetry forcing these cancellations. Applying the
naturalness criterion of 't Hooft  to the quark masses in the 2HDM
of type III we find that\cite{Crivellin:2012ye}
\begin{eqnarray}
|v_{u(d)} \epsilon^{d(u)}_{ij}|\leq \left|V_{ij}\right|\,{\rm max
}\left[m_{d_i(u_i)},m_{d_j(u_j)}\right]\,.
\end{eqnarray}
which leads to
\begin{eqnarray}
|\epsilon^{d(u)}_{ij}|\leq \frac{\left|V_{ij}\right|\,{\rm max
}\left[m_{d_i(u_i)},m_{d_j(u_j)}\right]}{|v_{u(d)}|}\,.\label{constr}
\end{eqnarray}
\begin{figure}
  \includegraphics[width=6.5cm]{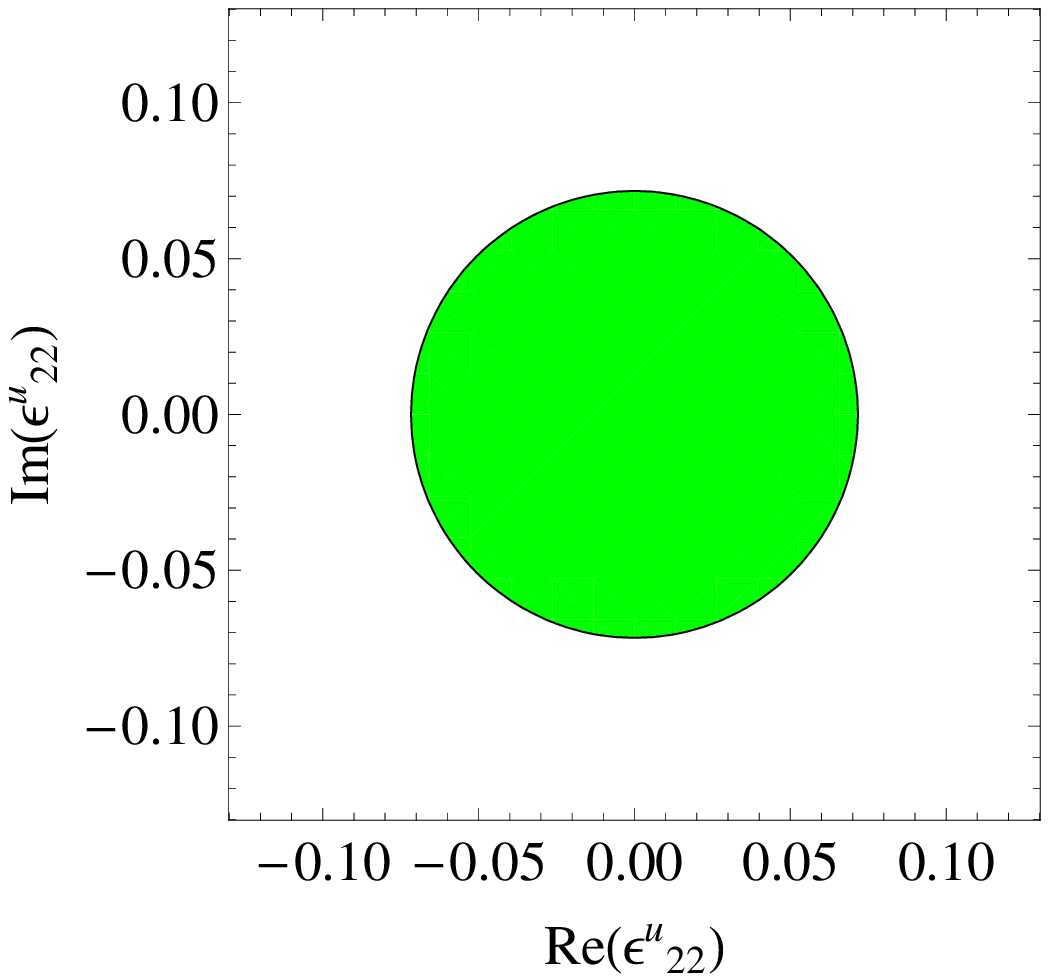}
  \includegraphics[width=6.5cm]{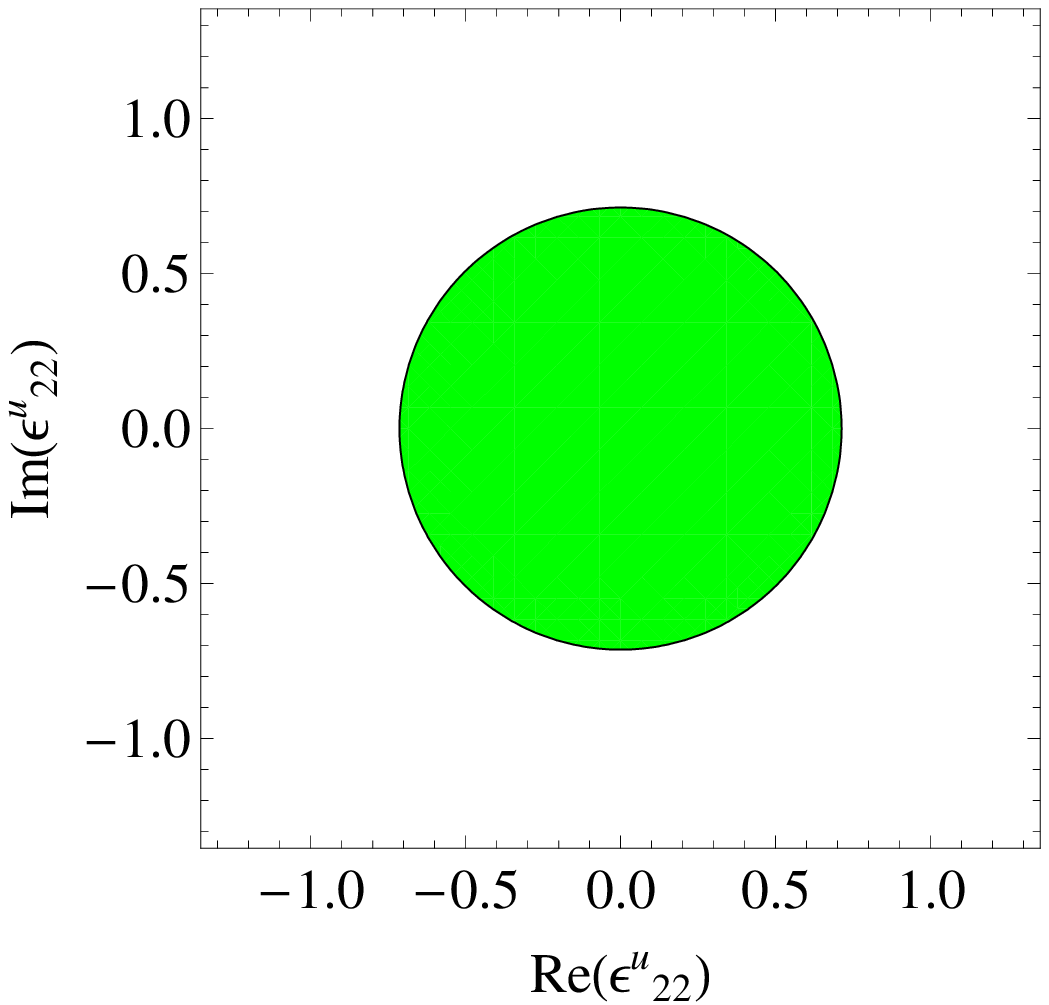}
  \caption{Constraints on  $\epsilon^u_{22}$. Left plot
corresponding to $\tan\beta =10$  while right plot corresponding
to $\tan\beta =100$.}\label{higsplane}
\end{figure}
Clearly from the previous equation that
$\epsilon^u_{11},\epsilon^d_{11},\epsilon^d_{22}$ will be severely
constrained by their small masses while $\epsilon^u_{22}$ will be
less constrained. Clearly from Eq.(\ref{constr}), the constraints
imposed  on $\epsilon^u_{22}$ are $\tan\beta$ dependent. We now
apply the constraints imposed on the real and imaginary parts of
$\epsilon^{u}_{22}$ corresponding to two different values of
$\tan\beta$ namely for two cases $\tan\beta =10$ and $\tan\beta
=100$ using Eq.(\ref{constr}). In Fig.(\ref{higsplane}) we show
the allowed regions for the two cases. Clearly the constraints are
sensitive to the value of $\tan\beta $ where the constraints are
weak for large values of $\tan\beta $. Since $C^H_1$ and $C^H_4$
 are proportional to $\epsilon^u_{22}$ thus they will be several
 order of magnitudes larger than $C^H_2$  and $C^H_3$. In fact this
 conclusion can be seen from Eq.(\ref{Higgsw}) and thus  in
 our analysis we  drop $C^H_2$  and $C^H_3$. Now possible other
 constraints on $\epsilon^u_{22}$ can be obtained from  $D-\bar{D}$ mixing, $K-\bar{K}$ mixing.
For $K-\bar{K}$ mixing, the new contribution from charged Higgs
mediation corresponding to top quark running in the loop will be
much dominant than the contribution in the case where the charm
quark runing in the loop. This is due to the dependency of the
contribution on the ratio of the quark mass running in the loop to
the charged Higgs mass. Thus the expected constraints from
$K-\bar{K}$ mixing might  be relevant  on $\epsilon^u_{32}$and
$\epsilon^u_{31}$ not on $\epsilon^u_{22}$. In fact, as mentioned
in ref.\cite{Crivellin:2012ye}, the constraints on
$\epsilon^u_{32}$and $\epsilon^u_{31}$ are even weak and
$\epsilon^u_{32}$and $\epsilon^u_{31}$ can be sizeable. By a
similar argument we can neither  use the process $b\rightarrow
s\gamma$ nor the Electric dipole moment (EDM) to constraint
$\epsilon^u_{22}$.
Regarding $D-\bar{D}$ mixing one expects a similar situation like
that in $K-\bar{K}$ about the dominance of top quark contribution.
However due to the CKM suppression factors the top quark
contribution will be smaller than the charm contribution.
\subsubsection{$D-\bar{D}$ mixing constraints}\label{DDbar}
We  take into accounts only box diagram that contribute to
$D-\bar{D}$ mixing  mediated by exchanging strange quark and
charged Higgs. Other contributions from box diagram mediated by
down or bottom quarks and charged Higgs are suppressed by the CKM
factors. Since SM contribution to $D-\bar{D}$ mixing is very small
we neglect its contribution and neglect its interference with
charged Higgs mediation contribution.  Thus  effective Hamiltonian
for this case can be written as:
\be {\mathcal H}^{|\Delta
C|=2}_{H^{\pm}}= \frac{1}{m^2_{H^{\pm}}}\sum^4_{i=1} C_i(\mu)
Q_i(\mu)+\tilde{C}_i(\mu) \tilde{Q}_i(\mu),
\ee where $C_i,\tilde{C}_i$ are the Wilson coefficients obtained by
perturbative QCD running from $M_H$ scale to the scale $\mu$
relevant for hadronic decay and $Q_i,\tilde{Q}_i$ are the relevant
local operators at low energy scale
\bea
Q_1 &=&(\bar{u}\gamma^{\mu} P_L c)(\bar{u} \gamma_{\mu} P_L c),\nonumber\\
Q_2 &=&(\bar{u} P_L c)(\bar{u} P_L c),\nonumber\\
Q_3 &=&(\bar{u}\gamma^{\mu} P_L c)(\bar{u} \gamma_{\mu} P_R c),\nonumber\\
Q_4 &=&(\bar{u} P_L c)(\bar{u} P_R c),\nonumber\\
\label{QDoper}
\eea
where we drop color indices and the operators $\tilde{Q}_i$ can be
obtained from $Q_i$ by changing the chirality $L\leftrightarrow
R$. The Wilson coefficients $C_i$,
are given by
\begin{eqnarray}
C_1 &=& \frac{ I_1(x_s)}{64 \pi^2 } \bigg(\sum\limits_{j = 1}^3
{\sin\beta\,V^*_{2j} \left( \frac{m_s}{v_d} \delta_{j2}-
\epsilon^{ d}_{j2}\tan\beta \right)}\bigg)^2\bigg(\sum\limits_{k =
1}^3 {\sin\beta\,V_{1k}
\left( \frac{m_s}{v_d}\delta_{k2}- \epsilon^{ d}_{k2}\tan\beta \right)}\bigg)^2,\nonumber\\
C_2 &=& \frac{m^2_s I_2(x_s)}{16 \pi^2 m^2_{H^{\pm}}}
\bigg(\sum\limits_{j = 1}^3 {\sin\beta\,V^*_{2j} \left(
\frac{m_s}{v_d} \delta_{j2}- \epsilon^{ d}_{j2}\tan\beta
\right)}\bigg)^2 \bigg(\sum\limits_{k = 1}^3 {\cos\beta\, V_{k2}
\left( \frac{m_u }{v_u} \delta_{k1}- \epsilon^{
u\star}_{k1}\tan\beta
\right)}\bigg)^2  ,\nonumber\\
 C_3 &=& \frac{ I_1(x_s)}{64 \pi^2 } \bigg(\sum\limits_{j = 1}^3 {\sin\beta\,V^*_{2j}  \left(
\frac{m_s}{v_d} \delta_{j2}- \epsilon^{ d}_{j2}\tan\beta
\right)}\bigg)\bigg(\sum\limits_{k = 1}^3 {\sin\beta\,V_{1k}
\left( \frac{m_s}{v_d}\delta_{k2}- \epsilon^{ d}_{k2}\tan\beta
\right)}\bigg)\nonumber\\&\times&\bigg(\sum\limits_{l = 1}^3
{\cos\beta\, V_{l2} \left( \frac{m_u }{v_u} \delta_{l1}-
\epsilon^{ u\star}_{l1}\tan\beta
\right)}\bigg)\bigg(\sum\limits_{n = 1}^3 {\cos\beta\, V^*_{n2}
\left( \frac{m_c }{v_u} \delta_{n2}-
\epsilon^{ u\star}_{n2}\tan\beta \right)}\bigg),\nonumber\\
C_4 &=&  \frac{m^2_s I_2(x_s)}{16 \pi^2 m^2_{H^{\pm}}}
\bigg(\sum\limits_{j = 1}^3 {\sin\beta\,V^*_{2j} \left(
\frac{m_s}{v_d} \delta_{j2}- \epsilon^{ d}_{j2}\tan\beta
\right)}\bigg)\bigg(\sum\limits_{k = 1}^3 {\sin\beta\,V_{1k}
\left( \frac{m_s}{v_d}\delta_{k2}- \epsilon^{ d}_{k2}\tan\beta
\right)}\bigg)\nonumber\\&\times&\bigg(\sum\limits_{l = 1}^3
{\cos\beta\, V_{l2} \left( \frac{m_u }{v_u} \delta_{l1}-
\epsilon^{ u\star}_{l1}\tan\beta
\right)}\bigg)\bigg(\sum\limits_{n = 1}^3 {\cos\beta\, V^*_{n2}
\left( \frac{m_c }{v_u} \delta_{n2}-
\epsilon^{ u\star}_{n2}\tan\beta \right)}\bigg).\nonumber\\
\label{DDWilson}
\end{eqnarray}
where $x_s=m^2_s/m^2_{H^{\pm}}$ and the integrals are defined as
follows:
\bea I_1 (x_s) &=& \frac{x_s+1 }{(x_s-1)^2}+\frac{-2 x_s\ln
(x_s)}{(x_s-1)^3},\nonumber\\
I_2(x_s)&=& \frac{-2 }{(x_s-1)^2}+\frac{(x_s+1)\ln
(x_s)}{(x_s-1)^3} \eea
 The Wilson coefficients  $\tilde{C}_i$  are
given by
\begin{eqnarray}
\tilde{C}_1 &=& \frac{ I_1(x_s)}{64 \pi^2 } \bigg(\sum\limits_{j =
1}^3 {\cos\beta\, V_{j2} \left( \frac{m_u }{v_u} \delta_{j1}-
\epsilon^{ u\star}_{j1}\tan\beta
\right)}\bigg)^2\bigg(\sum\limits_{k = 1}^3 {\cos\beta\, V^*_{k2}
\left( \frac{m_c }{v_u} \delta_{k2}-
\epsilon^{ u\star}_{k2}\tan\beta \right)}\bigg)^2, \nonumber\\
\tilde{C}_2 &=& \frac{m^2_s I_2(x_s)}{16 \pi^2 m^2_{H^{\pm}}}
\bigg(\sum\limits_{j = 1}^3 {\cos\beta\, V^*_{j2} \left( \frac{m_c
}{v_u} \delta_{j2}- \epsilon^{ u\star}_{j2}\tan\beta
\right)}\bigg)^2\bigg(\sum\limits_{k = 1}^3 {\sin\beta\,V_{1k}
\left( \frac{m_s}{v_d} \delta_{k2}- \epsilon^{ d}_{k2}\tan\beta
\right)} \bigg)^2  \nonumber\\
 \tilde{C}_3 &=& C_3 ,\nonumber\\
\tilde{C}_4 &=& C_4. \label{DDWilson}
\end{eqnarray}
 Our set of operators $Q_1$, $Q_2$ and  $Q_4$ given in Eq.(\ref{QDoper}) are
 equivalent to their corresponding operators  given in
Refs.\cite{Petrov:2010gy,Petrov:2011un} while the operators
$\tilde{Q}_1$ and $\tilde{Q}_2$ are  equivalent to $Q_6$ and $Q_7$
given in the same references respectively. Moreover $Q_3$, given
in Eq.(\ref{QDoper}), can be related to $Q_5$ in
Refs.\cite{Petrov:2010gy,Petrov:2011un} by Fierz identity. For the
rest of the operators,  $\tilde{Q}_3$ and  $\tilde{Q}_4$,  they
are equivalent to $Q_5$ and  $Q_4$ in
Refs.\cite{Petrov:2010gy,Petrov:2011un} since their matrix
elements are equal. Thus our Wilson coefficients can  be subjected
to the  constraints given in
Ref.\cite{Petrov:2010gy,Petrov:2011un} and so we find that
  \bea \mid C_{1} \mid &\leq& 5.7\times
10^{-7}\bigg[\frac{m_{H^{\pm}}}{1\, TeV}\bigg]^2\nonumber\\
 \mid C_{2} \mid &\leq& 1.6\times
 10^{-7}\bigg[\frac{m_{H^{\pm}}}{1\,
TeV}\bigg]^2\nonumber\\
\mid C_{3} \mid &\leq& 3.2\times
10^{-7}\bigg[\frac{m_{H^{\pm}}}{1\, TeV}\bigg]^2\nonumber\\
 \mid C_{4} \mid &\leq& 5.6\times
10^{-8}\bigg[\frac{m_{H^{\pm}}}{1\,
TeV}\bigg]^2\nonumber\\\label{DC1til}\eea
the constraints on $\tilde{C}_1-\tilde{C}_4$ are similar to those
$C_1-C_4$. As can be seen from Eq.(\ref{DC1til}) the constraints
on the Wilson coefficients will be strong for small charged Higgs
masses.
We can proceed now to derive  the constraints on $\epsilon^{
u}_{22}$ using the upper bound on $\tilde{C}_2$ for instance.
 Keeping terms corresponding to first order in $\lambda $ where $\lambda $
is the CKM parameter we find that, for $m _{H^{\pm}}=300$ GeV and
$\tan\beta=55$ \bea
 \tilde{C}_2\times 10^{12}  &\simeq& 3 \,\bigg(-53.6 \,\epsilon^{ d}_{12} - 12.7 \,\epsilon^{
 d}_{22}+0.007\bigg)^2
\bigg( -12.4 \, \epsilon^{ u\,*}_{12} -53.4 \,
\epsilon^{u\,*}_{22} +0.007 \bigg)^2\label{Ctild22}\eea
While for $m _{H^{\pm}}=300$ GeV and $\tan\beta=500$ we find
\bea
 \tilde{C}_2\times 10^{14}  &\simeq& 3.6 \,\bigg(-487.1 \,\epsilon^{ d}_{12} -115.0 \,\epsilon^{
 d}_{22}+0.06\bigg)^2
\bigg( -112.5\, \epsilon^{ u\,*}_{12} -486.7 \,
\epsilon^{u\,*}_{22} +0.007 \bigg)^2 \label{Ctild21}\eea
In both  Eqs.(\ref{Ctild21},\ref{Ctild22}) we can drop terms
proportional to $\epsilon^{ u\,*}_{12} $ to a good approximation
as they have small coefficients in comparison to $\epsilon^ u_{22}
$ and also since $\epsilon^{ u,d}_{ij} $ with $i\neq j $ are
always smaller than the diagonal elements $\epsilon^{ u,d}_{ii} $.
On the other hand we know that $\epsilon^d_{12} $ can not be large
to not allow flavor changing neutral currents and so we can drop
terms proportional to $\epsilon^d_{12}$ in
Eqs.(\ref{Ctild21},\ref{Ctild22}) to a good approximation also.
thus we are left with $\epsilon^d_{22}$  and $\epsilon^u_{22}$ in
both Eqs.(\ref{Ctild21},\ref{Ctild22}). Comparing their
coefficients shows that $\epsilon^u_{22}$ has a large coefficient
and thus we can drop $\epsilon^d_{22}$ terms.   An alternative way
is to  assume that $\epsilon^{ u}_{22}$ terms are the dominant
ones in comparison to the  other $\epsilon^{ u,d}_{ij}$ terms and
proceed to set upper bounds on $\epsilon^{ u}_{22}$.  In fact even
if we consider other Wilson coefficients rather than $\tilde{C}_2$
this conclusion will not be altered. Under the assumption
$\epsilon^{d}_{12} = \epsilon^{d}_{22}=\epsilon^{u}_{12}=0$  and
using the upper bound corresponding to $ m _{H^{\pm}}=300 $ GeV on
$ \tilde{C}_{2}$, using Eq.(\ref{DC1til}),  which reads in this
case  \bea \mid \tilde{C}_{2} \mid &\leq& 1.4 \times
10^{-8}\label{DC1ti2}\eea

Clearly from Eqs.(\ref{Ctild22},\ref{Ctild21},\ref{DC1ti2}) the
bounds that can be obtained on $\epsilon^{ u}_{22}$ will be so
loose and thus $D-\bar{D}$ mixing can not lead to a strong
constraints on $\epsilon^{ u}_{22}$.

\subsubsection{$D_q\to\tau\nu $ constraints}\label{DDbar}

The decay modes $D_q\to\tau\nu$ where $q=d$ or $q=s$ can be
generated in the SM  at tree level  via W boson  mediation. Within
the  2HDM of type III under consideration, the charged Higgs can
 mediate these decay modes at tree level also and hence the total
branching ratios, following a similar notations in
Ref.\cite{Crivellin:2012ye}, can be expressed as
\bea {\mathcal B}(D^+_q\to\tau^+\nu) & =&
\frac{G_F^2|V_{cq}|^2}{8\pi} m_\tau^2 f_{D_q}^2 m_{D_q}
\left(1-\frac{m_\tau^2}{m_{D_q}^2}\right)^2 \tau_{D_q} \nonumber  \\
&&\times
\left| 1+ \frac{m_{D_q }^{2}}{(m_c+m_q )\, m_{\tau}}
\frac{(C_{R}^{cq\,*}-C_L^{cq\,*})}{C_{SM}^{cq\,*}}   \right|^2\,.
\eea
Where we have used \cite{Na:2012uh}
\be \langle 0|\bar{q} \gamma^5 c |D_q\rangle= \frac{f_{D_q}m_{D_q
}^{2}}{(m_c+m_q )}\ee
Where the SM Wilson coefficient is given by $ C_{{\rm SM}}^{cq} =
{ 4 G_{F}} \; V^{}_{cq}/{\sqrt{2}}$  and the Wilson coefficients
$C_{L}^{cq}$ and $C_{R}^{cq}$ at the matching scale are given by
\begin{equation}
\renewcommand{\arraystretch}{1.5}
\begin{array}{l}
 C_{R(L)}^{cq} = \frac{{ -1}}{M_{H^{\pm}}^{2}} \; \Gamma_{cq}^{LR(RL),H^{\pm}} \; \frac{m_\tau}{v}\tan\beta  \,,
 \end{array}
\label{CRQ}\end{equation}
with the vacuum expectation value $v\approx174{\rm GeV}$ and
$\Gamma_{cq}^{LR(RL),H^{\pm}}$ can be read from Eq.(\ref{Higgsv}).
Setting the charged Higgs contribution to zero and $f_{D_s}=248\pm
2.5$ MeV \cite{Davies:2010ip}, we find that ${\mathcal
B}^{SM}(D^+_d\to\tau^+\nu) \simeq 9.5\times 10^{-4} $ and
${\mathcal B}^{SM}(D^+_s\to\tau^+\nu) = (5.11\pm 0.11)\times
10^{-2} $ which is in close agreement with the results in
Ref.\cite{Akeroyd:2007eh,Mahmoudi:2007vz,Mahmoudi:2008tp}. The
experimental values of these Branching ratios are given by
${\mathcal B}(D^+_d\to\tau^+\nu) < 2.1\times 10^{-3} $
\cite{Rubin:2006nt} while  ${\mathcal B}(D^+_s\to\tau^+\nu) =
(5.38\pm 0.32)\times 10^{-2} $\cite{Asner:2010qj}.  Keeping the
terms that are proportional to the dominant CKM elements we find
for $ q=d $ \bea {\Gamma_{c d }^{H^ \pm\,RL\,\rm{eff} } } &=&
{\cos\beta\,V_{11}} \left( -\epsilon^{ u\,^*}_{12}\tan\beta
\right)\nonumber\\
{\Gamma_{c d }^{H^ \pm\,LR\,\rm{eff} } } &=& {\sin\beta\,V_{11}
\left( \frac{m_d }{v_d} - \epsilon^{ d}_{11}\tan\beta
\right)}\nonumber\\
 \label{Higgsww}
\eea
While for $ q=s $   we find
\bea {\Gamma_{c s }^{H^ \pm\,RL\,\rm{eff} } } &=&
{\cos\beta\,V_{22}} \left(\frac{m_c}{v_u}  -\epsilon^{
u\,^*}_{22}\tan\beta
\right)\nonumber\\
{\Gamma_{c s }^{H^ \pm\,LR\,\rm{eff} } } &=& {\sin\beta\,V_{22}
\left( \frac{m_s }{v_d} - \epsilon^{ d}_{22}\tan\beta
\right)}\nonumber\\
 \label{Higgswww}
\eea
Clearly from the last two equations, we need to consider the decay
mode $D^+_s\to\tau^+\nu $ to constrain $\epsilon^{ u}_{22}$. For
$\tan \beta = 10$ we find that
\bea {\Gamma_{c s }^{H^ \pm\,RL\,\rm{eff} } }\times 10^{-3}
&\simeq& 0.71 - 968.6\, \epsilon^{ u}_{22}
\nonumber\\
{\Gamma_{c s }^{H^ \pm\,LR\,\rm{eff} } }\times 10^{-3} &\simeq&
5.3 - 9686.0 \, \epsilon^{ d}_{22}
 \label{Higgswww}
\eea
Clearly the coefficient of $\epsilon^{ d}_{22}$ is one order of
magnitude larger than $\epsilon^{ u}_{22}$ and for larger $\tan
\beta$ one expects to be larger than. However, $\epsilon^{
d}_{22}$ is severely constraint by naturalness criterion and thus
we expect that  the term proportional to $\epsilon^{ u}_{22}$ to
be larger and thus in our analysis we can drop $\epsilon^{
d}_{22}$ term and proceed to obtain the required constraints.
\begin{figure}[tbhp]
  \includegraphics[width=6.5cm]{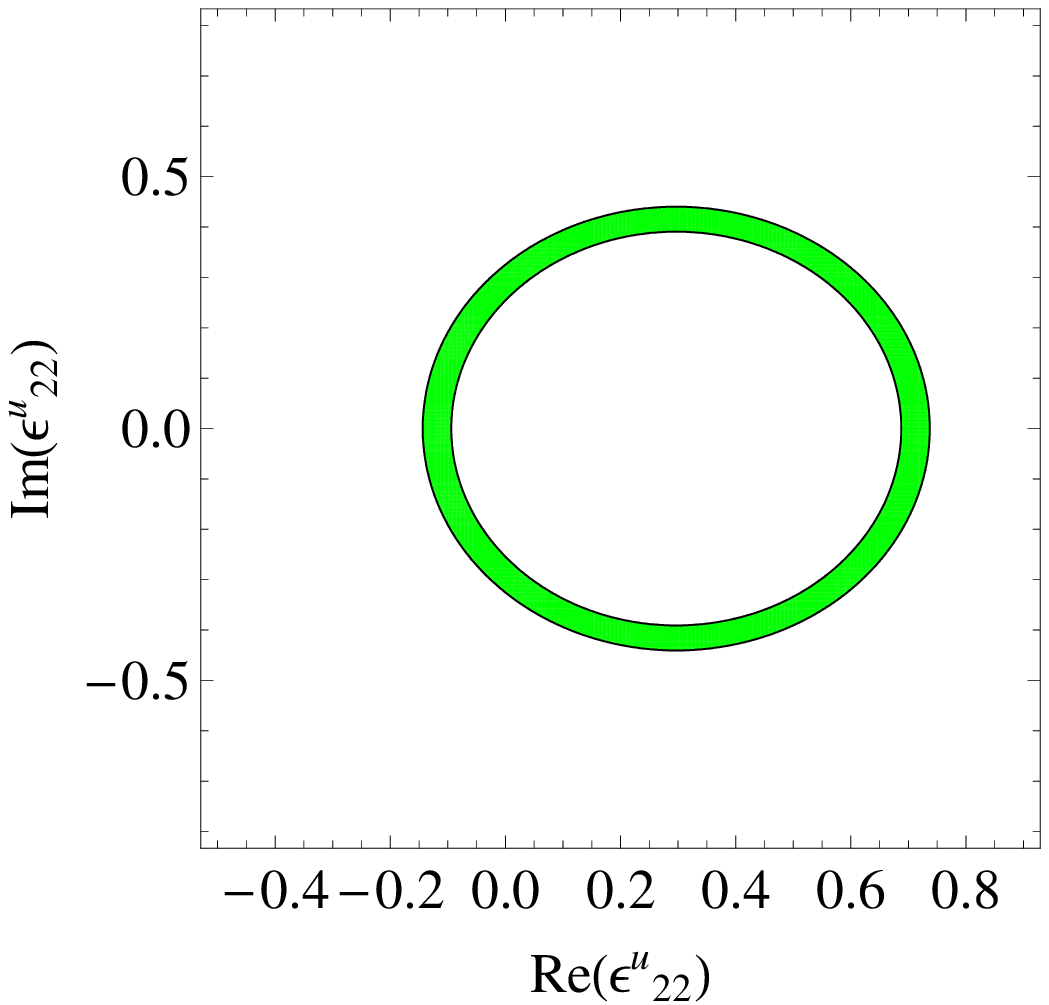}
  \includegraphics[width=6.5cm]{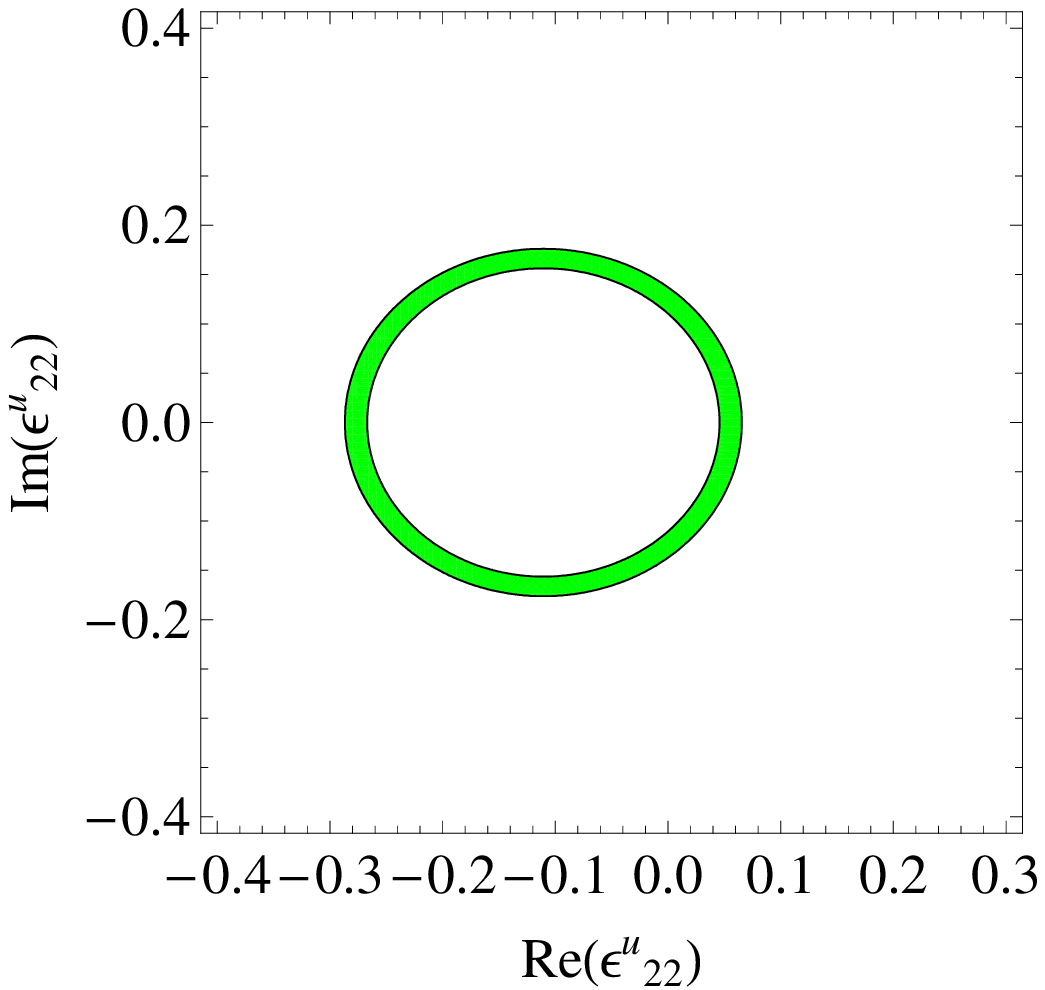}
  \caption{Constraints on  $\epsilon^u_{22}$ from ${\mathcal
B}(D^+_s\to\tau^+\nu)$. Left plot corresponding to $\tan\beta
=200$ while right plot corresponding to $\tan\beta =500$. In both
cases we take $m_{H^{\pm}}=200$ GeV.} \label{higsplaneDs1}\label{higsplaneDs1}
\end{figure}
\begin{figure}[tbhp]
  \includegraphics[width=6.5cm]{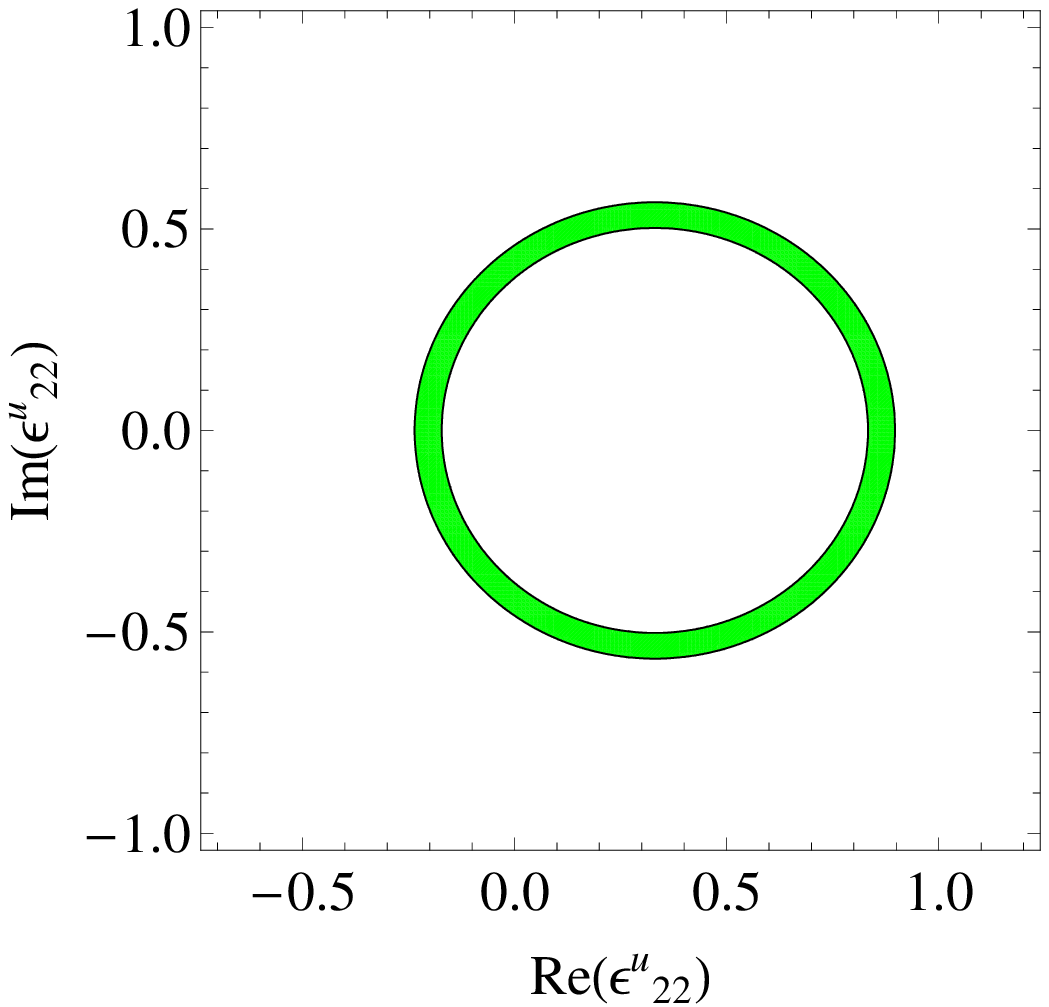}
  \includegraphics[width=6.5cm]{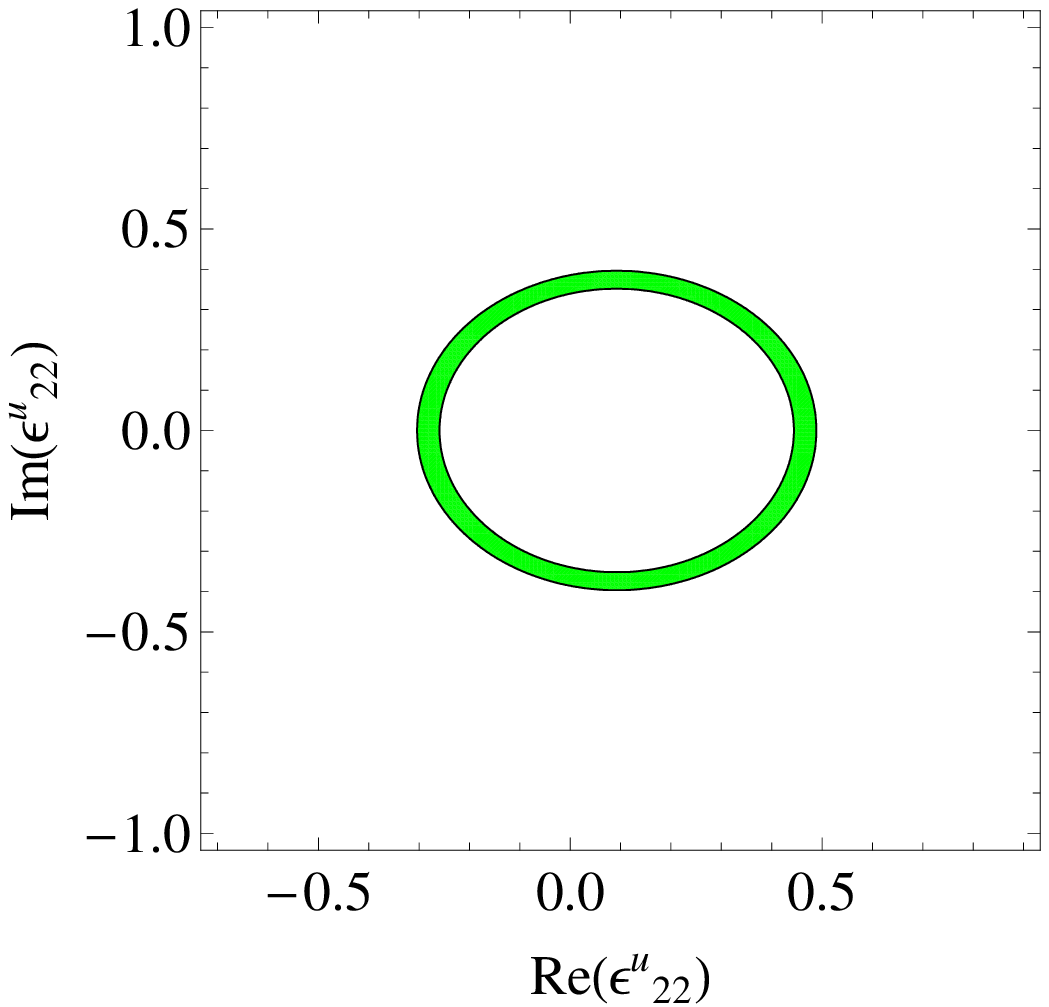}
  \caption{Constraints on  $\epsilon^u_{22}$ from ${\mathcal
B}(D^+_s\to\tau^+\nu)$. Left plot corresponding to $\tan\beta
=350$ while right plot corresponding to $\tan\beta =500$. In both
cases we take $m_{H^{\pm}}=300$ GeV.}\label{higsplaneDs2}
\end{figure}
We show in Figs.(\ref{higsplaneDs1},\ref{higsplaneDs2}) the
allowed regions for the real and imaginary parts of
$\epsilon^{u}_{22}$ corresponding to two different values of the
charged Higgs mass namely, $m_{H^{\pm}}=200$ and $m_{H^{\pm}}=300$
and for different values of $\tan\beta$. Our objective here is to
show the dependency of the constraints on $m_{H^{\pm}}$ and
$\tan\beta $. We see from the Figures that, for  $\tan\beta =500$,
the constraints  become loose with the increasing of
$m_{H^{\pm}}$. This is expected as Wilson coefficients of the
charged Higgs are inversely proportional to the square of
$m_{H^{\pm}}$ and thus their contributions to ${\mathcal
B}(D^+_s\to\tau^+\nu)$ become small for large $m_{H^{\pm}}$  which
in turn make the constraints obtained are loose. Another remark
from the figure is that the constraints  become strong  with the
increasing of the value of $\tan\beta$ which is expected also from
Eq.(\ref{CRQ}). This in contrast to the constraints derived by
applying the  naturalness criterion where we showed that the
constraints become loose with the increasing of the value of
$\tan\beta$.
\subsubsection{CP violation in Charged Higgs}
The total amplitude including SM and charged Higgs contribution
 can be written as
\be {\mathcal A}= \bigg(C^{SM}_1+\frac{1}{N} C^{SM}_2 +
\chi^{\pi^+}(C^H_1-C^H_4)\bigg)X_{D^0K^-}^{\pi^+}-
\bigg(C^{SM}_2+{1\over N} C^{SM}_1
+\frac{1}{2N}\big(C^H_1-\chi^{D^0}
C^H_4\big)\bigg)X_{K^-\pi^+}^{D^0}\label{HigsT}\ee with
$X^{P_1}_{P_2P_3}= if_{P_1}\Delta_{P_2P_3}^2
F_0^{P_2P_3}(m_{P_1}^2)$, $\Delta_{P_2P_3}^2=m_{P_2}^2-m_{P_3}^2$
  and $\chi^{\pi^+} $ and $\chi^{D^0}$ are previously defined as \bea \chi^{\pi^+}&=&{m_\pi^2\over (m_c-m_s)(m_u+m_d)}\nonumber\\
\chi^{D^0}&=& {m_D^2\over (m_c+m_u)(m_s-m_d)}\eea   The form of
the amplitude, ${\mathcal A}$,  shows how charged Higgs
contribution can affect only the short physics (Wilson
coefficients) without any new effect on the long range physics
(hadronic parameters). Thus strong phase will not be affected by
including charged Higgs contributions while the weak phase will be
affected. We can rewrite Eq.(\ref{HigsT}) in terms of the
amplitudes $T$ and $E$ introduced before in the case of the SM as
follows: \be {\mathcal A}= V^*_{cs}V_{ud}(T^{SM+H}+E^{SM+H})
\label{HigsTt}\ee where \bea T^{SM+H}= 3.14\times 10^{-6}&\simeq&
\frac{G_F}{\sqrt{2}}a^{SM+H}_1
f_{\pi}(m^2_D-m^2_K)F^{DK}_0(m^2_{\pi})\nonumber\\
E^{SM+H}= 1.53\times 10^{-6}e^{122^{\circ}i} &\simeq&
\frac{G_F}{\sqrt{2}}a^{SM+H}_2
f_D(m^2_K-m^2_{\pi})F^{K\pi}_0(m^2_D)\eea
where \bea a^{SM+H}_1&=&\bigg(C^{SM}_1+\frac{1}{N} C^{SM}_2 +
\chi^{\pi^+}(C^H_1-C^H_4)\bigg)\nonumber\\
&=&\bigg(a_1+\Delta  a_1+
\chi^{\pi^+}(C^H_1-C^H_4)\bigg)\label{a1t}\eea \bea a^{SM+H}_2=-
\bigg(a_2 +\Delta a_2+\frac{1}{2N}\big(C^H_1-\chi^{D^0}
C^H_4\big)\bigg)\label{a2t}\eea

The CP asymmetry can be obtained using the relation
\begin{eqnarray}
A_{CP} &=& {|{\mathcal A}|^2-|\bar {\mathcal A}|^2 \over
|{\mathcal A}|^2+|\bar {\mathcal A}|^2 } ={2|T^{SM+H}||
E^{SM+H}|\sin(\phi_1-\phi_2)\sin(-\alpha_E)\over |T^{SM+H} +
E^{SM+H} |^2 }
\end{eqnarray}
with  $\phi_i= Arg [a^{SM+H}_i] $ and $\alpha_E= Arg(\chi_E)$.  As an example  let us take
$Re(\epsilon^u_{22})=0.04$, $Im(\epsilon^u_{22})=0.03$ which is
allowed point for $\tan\beta =10$. In this case we find that for a
value of $m_{H^{\pm}}=500 $ GeV  we find that $ A_{CP}\simeq - 3.7
\times 10^{-5}$ while for $m_{H}=300 $ GeV we find that $
A_{CP}\simeq - 1 \times 10^{-4}$. Let us take another example
where $Re(\epsilon^u_{22})=-0.1$, $Im (\epsilon^u_{22})=-0.3$
which is allowed point for $\tan\beta =500$ and  $m_{H^{\pm}}=300
$ GeV. Repeating the same steps as above we find that $
A_{CP}\simeq 5.3\times 10^{-2}$. Clearly in charged Higgs models
the predicted CP asymmetry is so sensitive to the value of
$\tan\beta$ and to the value of Higgs mass.

\section{Conclusion}

In this paper, we have studied the Cabibbo favored non-leptonic
$D^0$ decays into $ K^- \pi^+$.  We have shown that  the Standard
Model prediction  for the corresponding CP asymmetry is strongly
suppressed and out of experimental range even taking into account
the large strong phases coming from the Final State Interactions.
Then we explored new physics models taking into account three
possible  extensions namely,  extra family, extra gauge bosons
within Left-Right Grand Unification models and extra Higgs Fields.
The fourth family model strongly improved SM prediction of the CP
asymmetry but still the  predicted CP asymmetry is far of the
reach of LHCB or SuperB factory as SuperKEKB. The most promising
models are no-manifest Left-Right extension of the SM where the LR
mixing between the gauge bosons permits us to get a strong
enhancement in the CP asymmetry. In such a model, it is possible
to get CP asymmetry of order $10 \%$ which is within the range of
LHCB and next generation of charm or B factory.  The
non-observation of such a huge CP asymmetry will strongly
constrain the parameters of this model. In multi Higgs extensions
of the SM, the 2HDM type III is the most attractive as it permits
to solve at the same time the puzzle coming from $B \to \tau \nu$
and give a large contribution to this CP asymmetry depending on
the charged Higgs masses and couplings. A maximal value of $ 5\%$
can be reached with a Higgs mass of $300$ GeV and large $tan
\beta$.

 \section*{Acknowledgements}
 G.F. thanks A. Crivellin for useful discussion. D. D. is  grateful to
Conacyt (M\'exico) S.N.I. and Conacyt project (CB-156618), DAIP
project (Guanajuato University) and PIFI (Secretaria de Educacion
Publica, M\'exico) for financial support. G.F. work is  supported
by research grants NSC 99- 2112-M-008- 003-MY3, NSC
100-2811-M-008-036 and NSC 101- 2811-M-008-022 of the National
Science Council of Taiwan.

\appendix
\section{Operators and other definitions}
We start by defining $X^{P_1}_{P_2P_3}$, where $P_i$ denotes a
pseudoscalar meson, as follows \be X^{P_1}_{P_2P_3}=
if_{P_1}\Delta_{P_2P_3}^2 F_0^{P_2P_3}(m_{P_1}^2)\label{ape1}\ee
where $\Delta_{P_2P_3}^2=m_{P_2}^2-m_{P_3}^2$. In terms of
$X^{P_1}_{P_2P_3}$ we find that
\begin{eqnarray}
<\pi^+|\bar u\gamma_\mu \gamma_5d |0><K^-|\bar s\gamma_\mu c|D^0>   &=& -X^{\pi^+}_{D^0K^-} \nonumber \\
<K^-\pi^+|\bar s\gamma_\mu d |0><0|\bar u\gamma_\mu  \gamma_5c|D^0> &=& X^{D^0}_{K^-\pi^+} \nonumber \\
<\pi^+|\bar u\gamma_5d |0><K^-|\bar s c|D^0> &=& -{m_\pi^2\over (m_c-m_s)(m_u+m_d)}X^{\pi^+}_{D^0K^-}\equiv -\chi^{\pi^+}X^{\pi^+}_{D^0K^-} \nonumber \\
<K^-\pi^+|\bar s d |0><0|\bar u \gamma_5c|D^0> &=& -{m_D^2\over
(m_c+m_u)(m_s-m_d)} X^ {D^0}_{K^-\pi^+}
 \equiv -\chi^{D^0} X^ {D^0}_{K^-\pi^+}\nonumber\\ \label{ape2}
\end{eqnarray}
 Using Eq.(\ref{ape2}) we get
\begin{eqnarray}
<K^-\pi^+|{\cal O}_1|D^0> &=& <K^-\pi^+| \bar s\gamma_\mu c_L\bar u\gamma_\mu d_L  |D^0>=<\pi^+|\bar u\gamma_\mu d_L |0><K^-|\bar s\gamma_\mu c_L|D^0>   \nonumber  \\
 &+& {1\over N} <K^-\pi^+| \bar s\gamma_\mu d_L |0><0|  \bar u\gamma_\mu c_L|D^0>=X_{D^0K^-}^{\pi^+}-{1\over N}X_{K^-\pi^+}^{D^0}
 \nonumber \\
<K^-\pi^+|{\cal O}_2|D^0> &=& <K^-\pi^+| \bar u\gamma_\mu c_L \bar
s\gamma_\mu d_L|D^0>=  <K^-\pi^+| \bar s\gamma_\mu d_L |0><0| \bar
u\gamma_\mu c_L|D^0>
\nonumber  \\
 &+& {1\over N}<\pi^+|\bar u\gamma_\mu d_L |0><K^-|\bar s\gamma_\mu c_L|D^0>=-X_{K^-\pi^+}^{D^0}+{1\over N}X_{D^0K^-}^{\pi^+}
 \nonumber \\
<K^-\pi^+| \bar s\gamma_\mu c_R\bar u\gamma_\mu d_R  |D^0> &=&
<\pi^+|\bar u\gamma_\mu d_R |0><K^-|\bar s\gamma_\mu c_R|D^0>
\nonumber  \\
 &+& {1\over N} <K^-\pi^+| \bar s\gamma_\mu d_R |0><0|  \bar u\gamma_\mu c_R|D^0>=-<K^-\pi^+|{\cal O}_1|D^0>
 \nonumber \\
 <K^-\pi^+| \bar u\gamma_\mu c_R  \bar s\gamma_\mu d_R|D^0> &=&   <K^-\pi^+| \bar s\gamma_\mu d_R |0><0|  \bar u\gamma_\mu c_R|D^0>
 \nonumber  \\
 &+& {1\over N}<\pi^+|\bar u\gamma_\mu d_R |0><K^-|\bar s\gamma_\mu c_R|D^0>=-<K^-\pi^+|{\cal O}_2|D^0>
 \nonumber \\
 <K^-\pi^+| \bar s\gamma_\mu c_L\bar u\gamma_\mu d_R  |D^0> &=& <\pi^+|\bar u\gamma_\mu d_R |0><K^-|\bar s\gamma_\mu c_L|D^0>
\nonumber  \\
 &-& {2\over N} <K^-\pi^+| \bar s d_{S+P} |0><0|  \bar u c_{S-P}|D^0>=-X^{\pi^+}_{D^0K^-}-{2\over N}\chi^{D^0} X^{D^0}_{K^-\pi^+}
 \nonumber \\
 <K^-\pi^+| \bar u\gamma_\mu c_L  \bar s\gamma_\mu d_R|D^0> &=&   <K^-\pi^+| \bar s\gamma_\mu d_R |0><0|  \bar u\gamma_\mu c_L|D^0>
 \nonumber  \\
 &-& {2\over N}<\pi^+|\bar ud_{S+P}|0><K^-|\bar sc_{S-P}|D^0>=-X^{D^0}_{K^-\pi^+}+{2\over N}\chi^{\pi^+}X^{\pi^+}_{D^0K^-}
 \nonumber \\
  <K^-\pi^+| \bar s\gamma_\mu c_R\bar u\gamma_\mu d_L  |D^0> &=& <\pi^+|\bar u\gamma_\mu d_L |0><K^-|\bar s\gamma_\mu c_R|D^0>
\nonumber  \\
 &-& {2\over N} <K^-\pi^+| \bar s d_{S-P} |0><0|  \bar u c_{S+P}|D^0>=X^{\pi^+}_{D^0K^-}+{2\over N}\chi^{D^0} X^{D^0}_{K^-\pi^+}
 \nonumber \\
 <K^-\pi^+| \bar u\gamma_\mu c_R  \bar s\gamma_\mu d_L|D^0> &=&   <K^-\pi^+| \bar s\gamma_\mu d_L |0><0|  \bar u\gamma_\mu c_R|D^0>
 \nonumber  \\
 &-& {2\over N}<\pi^+|\bar ud_{S-P}|0><K^-|\bar sc_{S+P}|D^0>=X^{D^0}_{K^-\pi^+}-{2\over N}\chi^{\pi^+}X^{\pi^+}_{D^0K^-} \nonumber \\
 \end{eqnarray}
and for the scalar ones
\begin{eqnarray}
<K^-\pi^+| \bar s c_L\bar u d_L  |D^0>
&=&\chi^{\pi^+}X_{D^0K^-}^{\pi^+}-{1\over
2N}\chi^{D^0}X_{K^-\pi^+}^{D^0}
 \nonumber \\
<K^-\pi^+| \bar u c_L  \bar s
d_L|D^0>&=&\chi^{D^0}X_{K^-\pi^+}^{D^0}-{1\over
2N}\chi^{\pi^+}X_{D^0K^-}^{\pi^+}
 \nonumber \\
<K^-\pi^+| \bar s c_R\bar u d_R  |D^0> &=&
-\chi^{\pi^+}X_{D^0K^-}^{\pi^+}+{1\over
2N}\chi^{D^0}X_{K^-\pi^+}^{D^0}
 \nonumber \\
 <K^-\pi^+| \bar u c_R  \bar s d_R|D^0> &=&-\chi^{D^0}X_{K^-\pi^+}^{D^0}+{1\over 2N}\chi^{\pi^+}X_{D^0K^-}^{\pi^+}
 \nonumber \\
 <K^-\pi^+| \bar s c_L\bar u d_R  |D^0> &=& -\chi^{\pi^+}X^{\pi^+}_{D^0K^-}+{1\over 2N} X^{D^0}_{K^-\pi^+}
 \nonumber \\
 <K^-\pi^+| \bar u c_L  \bar s d_R|D^0> &=& \chi^{D^0}X^{D^0}_{K^-\pi^+}+{1\over 2N}X^{\pi^+}_{D^0K^-}
 \nonumber \\
 <K^-\pi^+| \bar s c_R\bar u d_L  |D^0> &=& \chi^{\pi^+}X^{\pi^+}_{D^0K^-}-{1\over 2N} X^{D^0}_{K^-\pi^+}
 \nonumber \\
<K^-\pi^+| \bar u c_R  \bar s d_L|D^0> &=& -\chi^{D^0}X^{D^0}_{K^-\pi^+}-{1\over 2N}X^{\pi^+}_{D^0K^-} \nonumber \\
 \end{eqnarray}

where the Fierz's ordering has been used
\begin{eqnarray}
(\bar \psi_1\Psi_2)_L(\bar \psi_3\Psi_4)_L &=& (\bar \psi_1\Psi_4)_L(\bar \psi_3\Psi_2)_L,\hskip0.3cm (\bar \psi_1\Psi_2)_L(\bar \psi_3\Psi_4)_R = -2(\bar \psi_1\Psi_4)_{S+P} (\bar \psi_3\Psi_2)_{S-P} \nonumber \\
4\bar \psi_1 \psi_{2,\ S\pm P}\bar \psi_3\psi_{4,\ S\pm P} &=&  -2 \bar \psi_1 \psi_{4,\ S\pm P}\bar \psi_3 \psi_{2,\ S\pm P} -{1\over 2}\bar \psi_1(1 \pm \gamma_5)\sigma^{\mu\nu}\psi_4\bar \psi_3(1 \pm \gamma_5)\sigma_{\mu\nu}\psi_2  \nonumber  \\
2(T_a)_{\alpha \beta}(T_a)_{\gamma \delta} &=& \delta_{\alpha
\delta} \delta_{\beta \gamma}-{1\over N}\delta_{\alpha
\beta}\delta_{\gamma \delta},\hskip0.3cm
(T_a)_{\alpha \beta}(T_a)_{\gamma \delta} = {N_C^2-1\over 2N_C} \delta_{\alpha \delta} \delta_{\beta \gamma}-{1\over N_C}(T^a)_{\alpha \delta}(T^a)_{\beta \gamma}\nonumber \\
\end{eqnarray}

\end{document}